\PassOptionsToPackage{dvipsnames}{xcolor}
\documentclass[aip,jap,amsmath,amssymb,preprint,floatfix]{revtex4-2}

\usepackage{amsmath}
\usepackage{paralist}
\usepackage[utf8]{inputenc}
\usepackage{graphicx}
\usepackage{array}
\graphicspath{ {figures/} }
\usepackage[colorlinks=true, allcolors=blue]{hyperref}
\usepackage{svg}
\usepackage{multirow}
\svgpath{{figures/}}
\usepackage{rotating}
\usepackage{hyperref}
\usepackage{dcolumn}
\usepackage{bm}
\usepackage[T1]{fontenc}
\usepackage{mathptmx}

\usepackage{epstopdf}
\usepackage{multirow}
\usepackage{mathtools}
\usepackage{xcolor}
\usepackage{textcomp}
\usepackage{multirow}
\usepackage{tensor}

\renewcommand{\deg}{$^\circ$}

\begin{document}
\title{Selecting Alternative Metals for Advanced Interconnects}

\author{Jean-Philippe Souli\'e}
\affiliation{Imec, 3001 Leuven, Belgium}

\author{Kiroubanand Sankaran}
\affiliation{Imec, 3001 Leuven, Belgium}

\author{Benoit Van Troeye}
\affiliation{Imec, 3001 Leuven, Belgium}

\author{Alicja Le\'sniewska}
\affiliation{Imec, 3001 Leuven, Belgium}

\author{Olalla Varela Pedreira}
\affiliation{Imec, 3001 Leuven, Belgium}

\author{Herman Oprins}
\affiliation{Imec, 3001 Leuven, Belgium}

\author{Gilles Delie}
\affiliation{Imec, 3001 Leuven, Belgium}

\author{Claudia Fleischmann}
\affiliation{Imec, 3001 Leuven, Belgium}
\affiliation{KU Leuven, Department of Physics and Astronomy, 3001 Leuven, Belgium}

\author{Lizzie Boakes}
\affiliation{Imec, 3001 Leuven, Belgium}

\author{C\'edric Rolin}
\affiliation{Imec, 3001 Leuven, Belgium}

\author{Lars-\AA{}ke Ragnarsson}
\affiliation{Imec, 3001 Leuven, Belgium}

\author{Kristof Croes}
\affiliation{Imec, 3001 Leuven, Belgium}

\author{Seongho Park}
\affiliation{Imec, 3001 Leuven, Belgium}

\author{Johan Swerts}
\affiliation{Imec, 3001 Leuven, Belgium}

\author{Geoffrey Pourtois}
\affiliation{Imec, 3001 Leuven, Belgium}

\author{Zsolt T\H{o}kei}
\affiliation{Imec, 3001 Leuven, Belgium}

\author{Christoph Adelmann}
\email[Corresponding author. Email: ]{Christoph.Adelmann@imec.be}
\affiliation{Imec, 3001 Leuven, Belgium}

\begin{abstract}

Interconnect resistance and reliability have emerged as critical factors limiting the performance of advanced CMOS circuits. With the slowdown of transistor scaling, interconnect scaling has become the primary driver of continued circuit miniaturization. The associated scaling challenges for interconnects are expected to further intensify in future CMOS technology nodes. As interconnect dimensions approach the 10 nm scale, the limitations of conventional Cu dual-damascene metallization are becoming increasingly difficult to overcome, spurring over a decade of focused research into alternative metallization schemes. The selection of alternative metals is a highly complex process, requiring consideration of multiple criteria, including resistivity at reduced dimensions, reliability, thermal performance, process technology readiness, and sustainability. This tutorial introduces the fundamental criteria for benchmarking and selecting alternative metals and reviews the current state of the art in this field. It covers materials nearing adoption in high-volume manufacturing, materials currently under active research, and potential future directions for fundamental study. While early alternatives to Cu metallization have recently been introduced in commercial CMOS devices, the search for the optimal interconnect metal remains ongoing.

\end{abstract}

\maketitle

\clearpage

\section{Introduction}

Microelectronic circuits are central elements in myriads of electronic appliances in almost every aspect of today's life. Logic circuits, memory cells, and sensors are used to process, store, and detect information in diverse applications, ranging from computers and smartphones to automobiles and medical equipment. The success of microelectronics relies on the relentless miniaturization of the underlying building blocks, which, in the case of logic circuits based on transistors, has been epitomized by the famed Moore's law.\cite{moore_cramming_1965} Similar scaling trends also apply to other devices, for example for memory cells. The reduction of device dimensions in combination with the enormous increase of device density\cite{burg_moores_2021} has led to large performance benefits, but also to lower energy consumption per operation, and, at least for older generations, much reduced cost per function. For instance, the cost to fabricate one transistor has been reduced by a factor of $10^{9}$ since 1970.\cite{moore_11_2003}  

\begin{figure}
\includegraphics[width=0.85\textwidth]{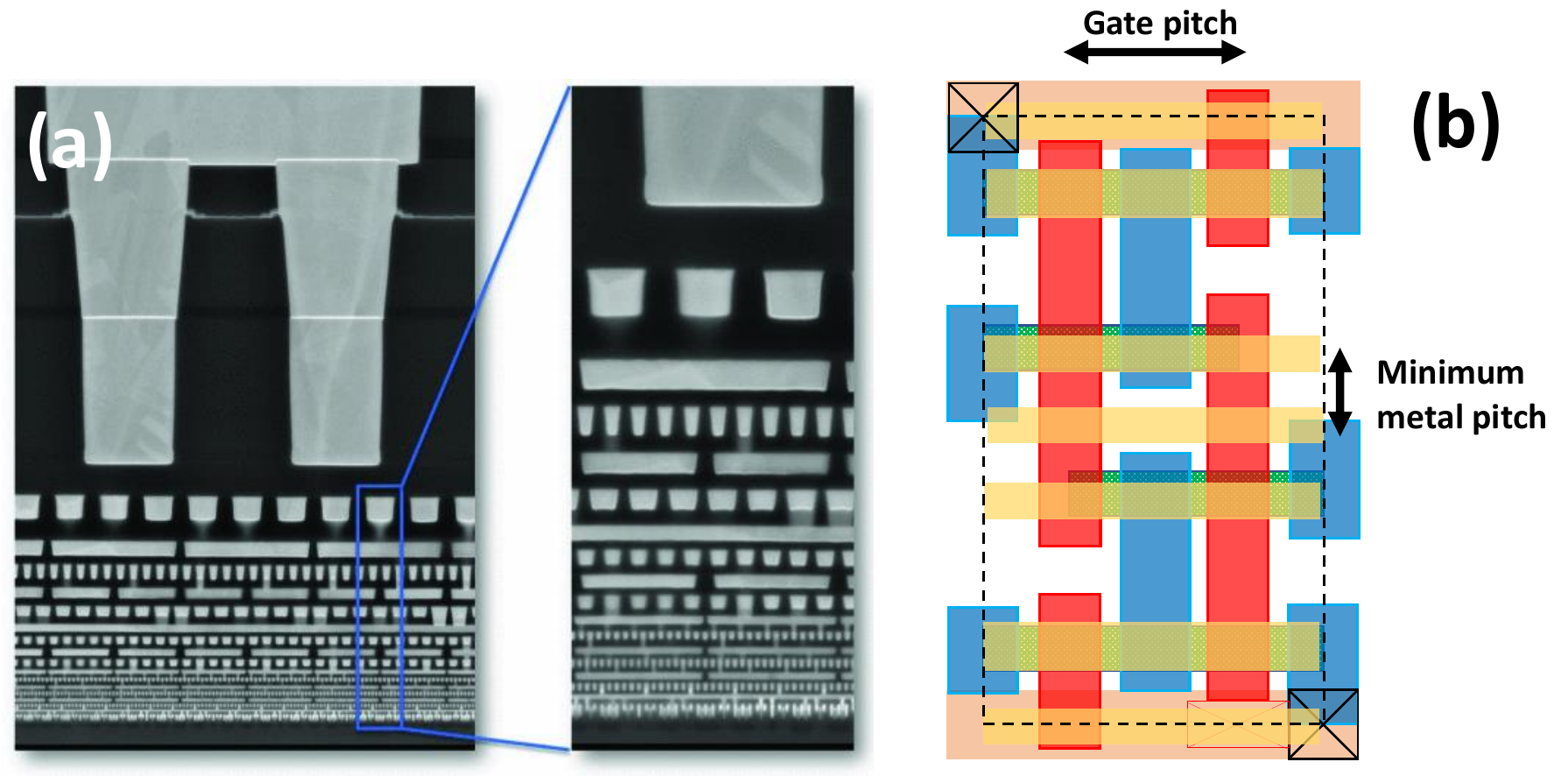} 
\caption{\label{IC_SRAM}(a) Cross-sectional transmission electron micrograph of the 16-level interconnect stack in Intel 4 technology. Reprinted with permission from Ref.~\onlinecite{sell_intel_2022}. (b) Layout of a static random-access memory (SRAM) cell in imec 3 nm technology, illustrating that the cell size is equally determined by transistor and metal pitch. For further details, see Refs.~\onlinecite{salahuddin_buried_2020,gupta_comprehensive_2021}.}
\end{figure}

In the public perception, Moore's law has been historically connected to scaling transistors or memory cells. However, scaling interconnects is of equal importance to uphold Moore's law. Interconnect lines and vias provide signal, power, and clock to the active components of the circuits, such as complementary metal--oxide--semiconductor (CMOS) transistors or memory elements, and thus are central in microelectronic circuits and systems with advanced functionality (Fig.~\ref{IC_SRAM}a). As an example, the area of a static random-access memory (SRAM) cell, used as cache memory in logic processors, is determined in one direction by the gate pitch (also termed ``contacted poly pitch'') of the transistors (the transistor size), but by the pitch of the metal lines (the interconnect pitch) in the orthogonal direction (Fig.~\ref{IC_SRAM}b). Hence, to reduce the cell area, both transistor and interconnect dimensions should be scaled. 

In earlier technology nodes, transistor performance traditionally improved as dimensions were scaled down, yielding the performance benefits described above. However, this trend does not extend to interconnects. Reducing the cross-sectional area of a wire inevitably increases its resistance per unit length, resulting in greater energy dissipation and higher resistive--capacitive delay ($RC$-delay). At current interconnect line dimensions, ranging from 12 to 15 nm (see Sec.~\ref{Sec:TechTargets}), interconnect performance has become a primary constraint on the overall performance of advanced microelectronic circuits.\cite{bohr_interconnect_1995, meindl_beyond_2003,ekekwe_power_2010, baklanov_advanced_2012, clarke_process_2014, hauschildt_advanced_2014, brain_interconnect_2016, vyas-chip_2018, bonilla_interconnect_2020, tokei_inflection_2020, li_interconnect_2022, nogami_interconnect_2023}

For transistors, scaling has been accompanied by significant architectural innovations.\cite{auth_22nm_2012, loubet_stacked_2017, jagannathan_vertical-transport_2021} In contrast, architectural changes in interconnects are limited, and thus performance enhancements mainly need to occur through materials and process innovation. One approach to reducing interconnect capacitance involves the use of dielectrics with lower permittivity (low-$\kappa$ dielectrics)\cite{baklanov_advanced_2012} or the introduction of air gaps.\cite{gosset_advanced_2005, chang_airgap_2023} However, these advancements have been hampered by the reduced mechanical stability of the resulting interconnect structures, leading to reliability issues during packaging.

\begin{figure}
\includegraphics[width=8cm]{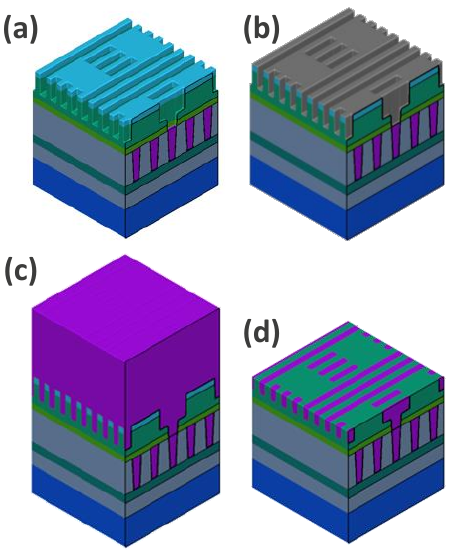} 
\caption{\label{DD_flow} Schematic of the Cu dual-damascene interconnect process integration flow: (a) via and trench patterning in low-$\kappa$ dielectric (green) using a hardmask (turquoise); (b) conformal barrier and liner deposition (grey); (c) Cu (over-)filling of vias and lines (purple); (d) chemical-mechanical polishing for line isolation and planarization, revealing Cu lines (purple) embedded in low-$\kappa$ dielectric (green).}
\end{figure}

Consequently, optimizing the metallization scheme has emerged as a critical focus of interconnect research in recent years.\cite{adelmann_alternative_2014,hauschildt_advanced_2014, zhang_ruthenium_2016, wen_ruthenium_2016, edelstein_20_2017, adelmann_alternative_2018, lin_future_2018, gall_resistivity_2020, gall_search_2020, gall_materials_2021, adelmann_intermetallic_2023, nogami_interconnect_2023} The current Cu dual-damascene metallization scheme (Fig.~\ref{DD_flow}), which replaced Al-based metallization post-1999\cite{edelstein_full_1997,kriz_overview_2008,gupta_copper_2009, baklanov_advanced_2012} is facing growing issues for several reasons. First, Cu requires diffusion barriers and adhesion liners to ensure the interconnect reliability. Without diffusion barriers (typically TaN-based), Cu migration into surrounding dielectrics leads to dielectric breakdown and shorting between adjacent lines (see Sec.~\ref{Sec:Reliability} ). Moreover, Cu electromigration becomes an increasing issue at scaled dimensions (see Sec.~\ref{Sec:Reliability}). This is mitigated by incorporating adhesion liner layers (typically Co) between the TaN barrier and Cu as well as by capping layers.\cite{gupta_copper_2009, baklanov_advanced_2012, oates_strategies_2014, hu_future_2018} However, the combined thickness of barrier and liner layers cannot be reduced below 2 to 3 nm without compromising functionality. Hence, in narrow lines, the high-resistivity barrier and liner layers occupy an increasingly significant fraction of the total metallization volume, reducing the available space for Cu, while contributing minimally to the wire's conductance. 

Furthermore, as elucidated in Sec.~\ref{Sec:Resistance}, the resistivity of Cu increases sharply at reduced dimensions due to the more pronounced effects of grain boundary and surface scattering. Both the increased resistivity and the decreasing volume fraction of Cu contribute to a rapid increase in the line and via resistances per unit length as interconnect dimensions are scaled down. This results in a significant deterioration of interconnect performance, even for relatively short lines. Additionally, the dual-damascene metallization integration process necessitates increasingly disruptive modifications to ensure void- and defect-free interconnects with robust mechanical stability.

\begin{figure}
\includegraphics[width=0.90\textwidth]{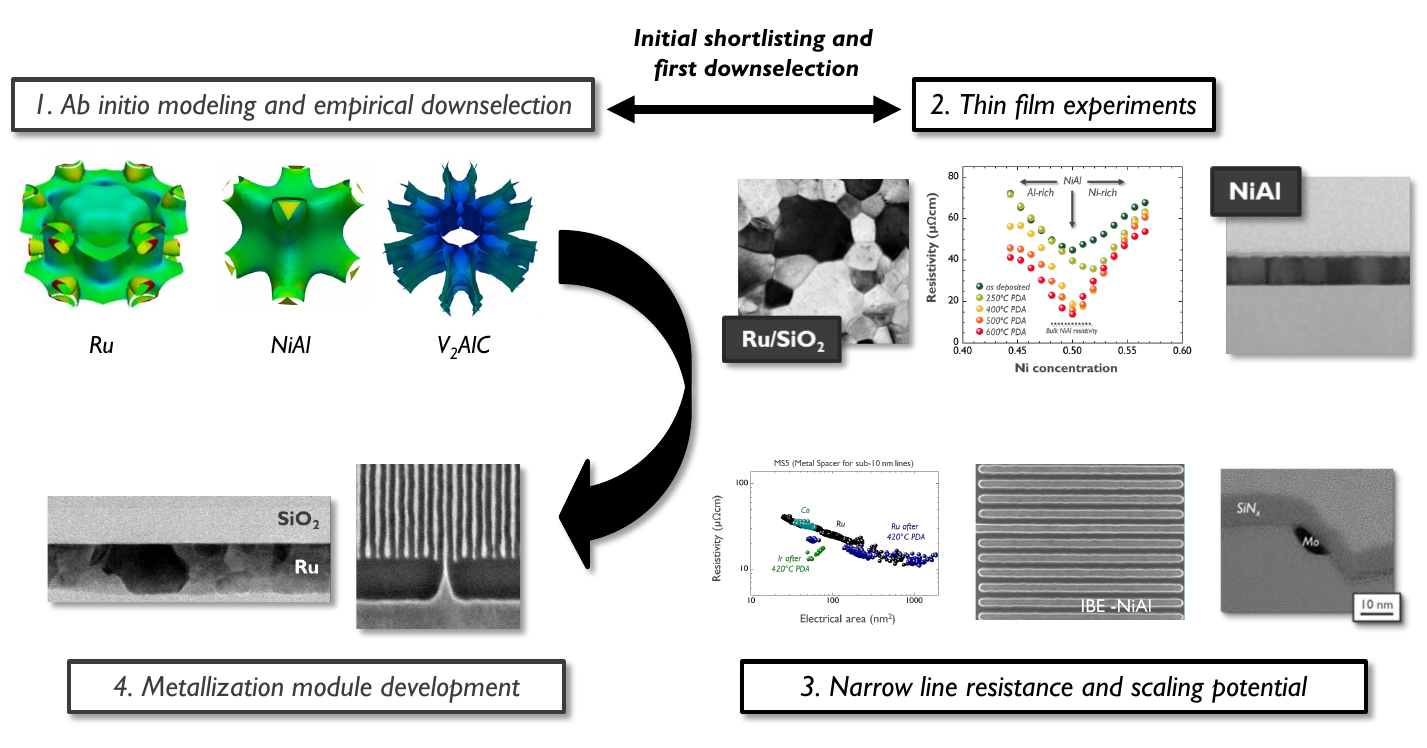} 
\caption{Imec workflow for identifying, downselecting, and benchmarking alternative interconnect metals: \textit{Ab initio} simulations are employed to identify potential candidate metals, which are subsequently downselected using a combination of thin film and nanowire experiments. The most promising materials are then selected for process module development and integration into scaled interconnects, followed by technology benchmarking.\label{fig:workflow}}
\end{figure}

These issues can be mitigated by selecting alternative metals that ideally do not require barrier and liner layers while exhibiting lower sensitivity of resistivity to nanoscale dimensions. Although this approach cannot reverse the increase in line resistance per unit length, we will demonstrate below that alternative metals and metallization schemes can outperform Cu at sufficiently small line widths. In this tutorial, we will discuss the various aspects relevant for the selection of potential alternative metals for advanced interconnects. The selection process is multifaceted and must address the challenge from different angles. To this end, we have developed a multistage framework to identify, downselect, and benchmark alternative metals for interconnect applications (Fig.~\ref{fig:workflow}).

The tutorial is organized as follows. We first examine the sensitivity of resistivity to nanoscale dimensions and introduce a material screening process. Next, we discuss reliability aspects, focusing on time-dependent dielectric breakdown as well as on electromigration. We then apply this selection process to elemental, binary, and ternary metals, highlighting current and future research directions. Given the strong connection between metal selection and future integration schemes, we will also briefly discuss integration and process considerations for alternative metals in upcoming technology nodes. Finally, with sustainability becoming increasingly crucial to minimizing the ecological footprint of the microelectronics industry, we introduce a life cycle assessment framework for interconnect metals and apply it to various promising candidates identified through our selection criteria.

\subsection{\label{Sec:TechTargets}Technology targets for future interconnects}

Historically, interconnect scaling has been guided by industry roadmaps such as the International Technology Roadmap for Semiconductors (ITRS),\cite{noauthor_itrs_2015} presently called International Roadmap for Devices and Systems (ITRDS).\cite{noauthor_itrs_nodate}   However, today, no industry-wide roadmap exists and also technology node nomenclature has become ambiguous. At present, commercial microelectronic chips feature minimum interconnect pitches around 25 nm, with further scaling anticipated to reach sub-20 nm metal pitches in the near future (see Tab.~\ref{Tab:Roadmap}). This implies that line widths, which are equivalent to half the metal pitch, will soon reach sub-10 nm dimensions. This is especially significant as the contacted poly pitch (transistor gate pitch) has become close to physical limits and is not expected to be scaled much further. Therefore, future area gains for CMOS circuits must primarily stem from transistor architecture innovation (\textit{e.g.} CFET) as well as continued interconnect pitch scaling.

\begin{table}
\caption{\label{Tab:Roadmap}Roadmap for interconnect dimensions (minimum metal pitch) in logic circuits. Adapted from Ref.~\onlinecite{Yamamoto_IEDM_Shortcourse}. For the current state of the art, see Refs.~\onlinecite{chang_critical_2022,wu_3nm_2022}. (HVM = high volume manufacturing; GAA = gate all-around; CFET = complementary field-effect transistor).\\}
    \centering
    \begin{tabular}{lcccccc}
\toprule
Year of HVM & 2024/25 & 2027/28 & 2029 & 2031 & 2033 & 2035 \\
\hline
Technology node & 2nm & 14A & 10A & 7A & 5A & 3A\\
Transistor technology & GAA & GAA & CFET & CFET & CFET & 2D\\
Min. metal pitch (nm) & 23 & 20 & 18 & 16 & 14 & 12\\
Gate pitch (nm) & 45 & 42 & 42 & 39 & 39& 36\\
\toprule
    \end{tabular}
\end{table}

As previously noted, such small line widths may not be compatible with the Cu dual-damascene metallization scheme employed today in scaled interconnects (Fig.~\ref{DD_flow}). While ongoing efforts to optimize Cu dual-damascene processing may yield incremental improvements, achieving the minimum interconnect dimensions in Tab.~\ref{Tab:Roadmap} will require disruptive approaches, potentially involving novel materials, processes, and integration schemes. As indicated in the table, the target critical dimensions for interconnect lines and vias, equivalent to half the metal pitch, fall within the range between 5 and 10 nanometers, providing guidelines for both metal selection and process development.

\section{Size Effects on the metal resistivity at nanoscale dimensions\label{Sec:Resistance}}

For decades, it has been well-established that the resistivity of metallic nanostructures, such as thin films and nanowires, is typically much higher than that of their bulk counterparts.\cite{fuchs_conductivity_1938, sondheimer_mean_1952, soffer_statistical_1967, mayadas_electrical_1969, mayadas_electrical-resistivity_1970, sambles_electrical_1980} This presents a significant challenge for interconnect scaling, as the increase in line and via resistance occurs at a much faster rate than what would be predicted solely by the reduction in geometrical dimensions, particularly for critical dimensions below 10 nm.

The resistivity increase can be primarily attributed to the combined effects of scattering at rough surfaces or interfaces\cite{fuchs_conductivity_1938, sondheimer_mean_1952} and scattering at grain boundaries.\cite{mayadas_electrical_1969,mayadas_electrical-resistivity_1970} To quantitatively describe this behavior, several transport models have been developed for thin films, accounting for top and bottom surface/interface scattering.\cite{mayadas_electrical-resistivity_1970,namba_resistivity_1970, marom_effect_2006, timoshevskii_influence_2008, feldman_dependence_2008, ke_resistivity_2009, rickman_resistivity_2012, moors_resistivity_2014,moors_modeling_2015,zhou_resistivity_2018,zhou_electrical_2018} However, it should be noted that no comprehensive one-dimensional transport model currently exists to describe nanowires, which are bounded by four surrounding surfaces. Despite this, the qualitative behavior of nanowires is expected to be similar to that of thin films, and thus, ``thin-film-derived'' resistivity models have been applied to understand the scaling behavior of nanowires as well.\cite{hinode_resistivity_2001,steinhogl_size-dependent_2002,maitrejean_experimental_2006, marom_size_dependent_2006, khoo_aspect_2007, graham_resistivity_2010, kim_structural_2011,chawla_electron_2011,smith_evaluation_2019}

In this section, we introduce the basic physics governing the resistivity of metal nanostructures. We will demonstrate that the mean free path of charge carriers in bulk metals is a critical parameter to characterize how resistivity depends on nanostructure dimensions. Following this, we discuss additional scattering mechanisms present in compound metals. Finally, we introduce \textit{ab initio} screening methods to calculate the mean free path and identify promising metals that may exhibit lower resistivity than Cu at the nanoscale.\cite{gall_electron_2016,dutta_thickness_2017,moors_first-principles-based_2022}

\subsection{Electron transport in metallic nanostructures}

\subsubsection{Semiclassical description of electron transport\label{Sec:Semiclassical}}

The most widely used approach to understand the increase in resistivity at reduced dimensions is based on the Boltzmann transport equation. The Boltzmann transport equation describes the statistical distribution of charge carriers in non-equilibrium conditions, \textit{e.g.}, when an electric field is applied.

In absence of temperature and composition gradients, as well as magnetic fields, the Boltzmann transport equation can be written as

\begin{equation}
-e \frac{\partial f^0_{n\mathbf{k}}}{\partial \epsilon_{n\mathbf{k}}} \mathbf{v}_{n\mathbf{k}}\cdot\mathbf{E} =  -\left.\frac{\partial g_{n\mathbf{k}}}{\partial t}\right|_{\text{scattering}} + \mathbf{v}_{n\mathbf{k}}\cdot\nabla_r g_{n\mathbf{k}}\, , \label{eq:BTE}
\end{equation}

\noindent where $e$ is the electron charge, $f^0_{n\mathbf{k}}$ the equilibrium Fermi--Dirac distribution for electrons of mode $n$, and $\mathbf{k}$ the wavevector of the electron with energy $\epsilon_{n\mathbf{k}}$ and velocity $\mathbf{v}_{n\mathbf{k}}$. $\mathbf{E}$ represents the external electric field, and $g_{n\mathbf{k}} = f_{n\mathbf{k}} - f^0_{n\mathbf{k}}$ denotes the deviation of the non-equilibrium electron distribution from the Fermi--Dirac equilibrium. $\left.\partial g_{n\mathbf{k}}/\partial t\right|_\mathrm{scattering}$ accounts for charge carrier scattering. 

In the linearized (semiclassical) approximation, referred to as the relaxation time approximation, the scattering term can be expressed as $\left. \partial g_{n\mathbf{k}}/\partial t \right|_\mathrm{scattering} = -g_{n\mathbf{k}}/\tau_{n\mathbf{k}}$, where $\tau_{n\mathbf{k}}$ represents the relaxation time. For high-purity bulk metals near room temperature, electron--phonon scattering dominates, with the corresponding relaxation time $\tau^\mathrm{ep}_{n\mathbf{k}}$. In polycrystalline films, grain boundaries introduce additional charge carrier scattering, characterized by the relaxation time $\tau^\mathrm{gb}_{n\mathbf{k}}$. When grain sizes are sufficiently large (indicating low disorder), both relaxation times can be considered independent, and the total relaxation time follows Matthiessen's rule

\begin{equation}
    (\tau_{n\mathbf{k}})^{-1} = (\tau^\mathrm{gb}_{n\mathbf{k}})^{-1} + (\tau^\mathrm{ep}_{n\mathbf{k}})^{-1}. \label{eq:matthiessen}
\end{equation}

For a bulk material with no spatial variation, the conductivity tensor $\underline{\underline{\sigma}}$ and resistivity tensor $\underline{\underline{\rho}}$ can be derived by noting that the current density is given by

\begin{equation}
    \mathbf{j} = \frac{-e}{A} \sum_n \int \int  g_{n\mathbf{k}} \mathbf{v}_{n\mathbf{k}}\mathrm{d}\mathbf{A} \, \mathrm{d}\mathbf{k}= \frac{-e^2}{N_\mathbf{k}\Omega}  \sum_{n\mathbf{k}}   \tau_{n\mathbf{k}} \frac{\partial f^0_{n\mathbf{k}}}{\partial \epsilon_{n\mathbf{k}}} \mathbf{v}_{n\mathbf{k}} \otimes \mathbf{v}_{n\mathbf{k}} \mathbf{E} = \underline{\underline{\sigma}} \mathbf{E} = \underline{\underline{\rho}}^{-1} \mathbf{E}\, , \label{eq:sigma}
\end{equation}

\noindent where $A$ represents the cross-sectional area through which electrons flow, $N_\mathbf{k}$ is the number of $\mathbf{k}$-points used to sample the Brillouin zone, and $\Omega$ denotes the volume of the unit cell.

Thus far, we have neglected the impact of the spatial gradient in Eq.~\eqref{eq:BTE}, which becomes relevant when considering a finite-size material, such as a thin film with thickness $h$. Due to the linear nature of the Boltzmann transport equation in the relaxation time approximation, the general solution is a superposition of a particular solution that accounts for boundary conditions and the solution derived for an infinitely large material. The particular solution is determined by the scattering behavior of electrons at the material's boundaries, \textit{i.e.}, at the top and bottom surfaces or interfaces of a thin film. This effect is further amplified in nanowires, where electrons can also scatter at lateral boundaries. Surface scattering can be either specular, when the electron preserves its total momentum and only the out-of-plane momentum component changes sign, or diffusive, when the electron's momentum is completely randomized. In practice, this results in the renormalization of the relaxation time for a thin film of thickness $h$, as described by\cite{VanTroeye2023}

\begin{equation}
\frac{\tau_{n\mathbf{k}}(h,\hat{u})}{\tau^{0}_{n\mathbf{k}}} =  1- (1-p)\frac{|\boldsymbol{\lambda}_{n\mathbf{k}}.\hat{u}|}{h} \frac{1-e^{-h/|\boldsymbol{\lambda}_{n\mathbf{k}}.\hat{u}|}}{1-p e^{-h/|\boldsymbol{\lambda}_{n\mathbf{k}}.\hat{u}|}}\, ,
\label{eq:rhoh}\end{equation}

\noindent where $\hat{u}$ denotes the unit vector normal to the thin film surface. $p$ is the reflection coefficient with $p=1$ representing purely specular surfaces and $p=0$ representing fully diffusive surfaces, and $\boldsymbol{\lambda}_{n\mathbf{k}} = \tau_{n\mathbf{k}} \mathbf{v}_{n\mathbf{k}}$ represents the mean free path of the charge carriers in the bulk metal.

From Eq.~\eqref{eq:rhoh}, it is evident that no resistivity increase occurs with decreasing film thickness $h$ for purely specular surfaces, whereas diffusive surfaces result in a strong dependence on $h$. For a metallic thin film with an isotropic Fermi surface, neglecting grain boundary scattering and assuming only electrons at the Fermi energy contribute, the conductivity can be described by the Fuchs--Sondheimer equation\cite{fuchs_conductivity_1938,sondheimer_mean_1952,sondheimer_mean_2001}

\begin{equation}
\frac{\sigma}{\sigma_0} = 1 - \frac{3\lambda}{2h}(1-p) \int_1^\infty \left(\frac{1}{t^3}-\frac{1}{t^5}\right) \frac{1-e^{-ht/\lambda}}{1-pe^{-h t/\lambda}} dt\, , \label{eq:fuchs}
\end{equation}

\noindent where $\sigma_0$ is the bulk conductivity and $\lambda$ the magnitude of the bulk mean free path.\cite{fuchs_conductivity_1938} Mayadas and Shatzkes subsequently extended this model to incorporate grain boundary scattering contributions,\cite{mayadas_electrical_1969} which are modeled as a series of partially transparent planes oriented perpendicular to the transport direction $x$. The distance between neighboring planes follows a normal distribution with average $d$ and variance $s$. In the regime of large variance $s$ for the distance between grain boundaries (more precisely the linear intercept length, related to grain size\cite{abrams_grain_1971,thorvaldsen_intercept_1997}), the relaxation time due to grain boundary scattering can be approximated by~\cite{mayadas_electrical_1969,mayadas_electrical-resistivity_1970}

\begin{equation}
(\tau^\mathrm{gb}_{\mathbf{k}})^{-1} = \frac{ v_{\mathbf{k}}^2}{d v_{\mathbf{k},x}}\frac{2R}{1-R} \, ,
\end{equation}

\noindent where $v_{\mathbf{k}}$ is the magnitude of the velocity and $v_{\mathbf{k},x}$ its component along the $x$-direction. This expression for the relaxation time can be combined with the bulk electron--phonon relaxation time using Matthiessen's rule, Eq.~\eqref{eq:matthiessen}, and the Fuchs--Sondheimer theory, Eq.~\eqref{eq:fuchs}, to derive the thickness- and grain-size dependent resistivity in thin films, given by\cite{mayadas_electrical-resistivity_1970}

\begin{equation}
\label{Eq:MS}\rho_\mathrm{tf} = \left[ \frac{1}{\rho_\mathrm{gb}} - \frac{6}{\pi \kappa \rho_0}\left( 1-p\right)\int\limits_0^{\pi/2} d\phi \int\limits_1^\infty dt \frac{\cos^2\phi}{H^2} \times\right. \left. \left(\frac{1}{t^3}-\frac{1}{t^5}\right) \frac{1-e^{-\kappa tH}}{1-pe^{-\kappa tH}} \right] ^{-1}\, ,
\end{equation}

\noindent with $\rho_\mathrm{gb} = \rho_0 \left[1-3\alpha /2 + 3\alpha^2 -3\alpha^3\ln\left( 1 + 1/\alpha\right) \right]^{-1}$, $H = 1 + \alpha /\cos\phi\sqrt{\left( 1-1/t^2\right)}$, $\kappa = h/\lambda$, and $\alpha = \left(\lambda /d\right) \times 2R\left(1-R\right)^{-1}$.

The derivation of Eq.~\eqref{Eq:MS} relies on the applicability of Matthiessen's rule for phonon and grain boundary scattering. However, this assumption may break down in nanocrystalline, nearly amorphous films with large disorder.\cite{mooij_electrical_1973,gurvitch_ioffe-regel_1981,tsuei_nonuniversality_1986} In such films, localization effects become prominent,\cite{imry_possible_1980,lee_disordered_1985} typically resulting in a notably reduced temperature coefficient of resistivity compared to bulk materials,\cite{mooij_electrical_1973,tsuei_nonuniversality_1986} as well as in different transport mechanisms, which cannot be described by Eq.~\eqref{Eq:MS}.

By contrast, the derivation of Eq.~\eqref{Eq:MS} does not require Matthiessen's rule to apply for surface scattering. In fact, Matthiessen's rule is typically not cromulent for surface scattering and grain boundary scattering in thin metallic films due to the renormalization of the mean free path by grain boundary scattering.\cite{mayadas_electrical-resistivity_1970}  Consequently, the individual contributions of bulk, surface, and grain boundary scattering to the thin film resistivity can be challenging to quantify.\cite{dutta_thickness_2017} Therefore, approximate versions of Eq.~\eqref{Eq:MS} that separate these contributions should be used with caution.

\subsubsection{Material-dependent scaling of thin film resistivity\label{Sec:Res_Mat_Dep}}

\begin{figure}
\includegraphics[width=0.95\textwidth]{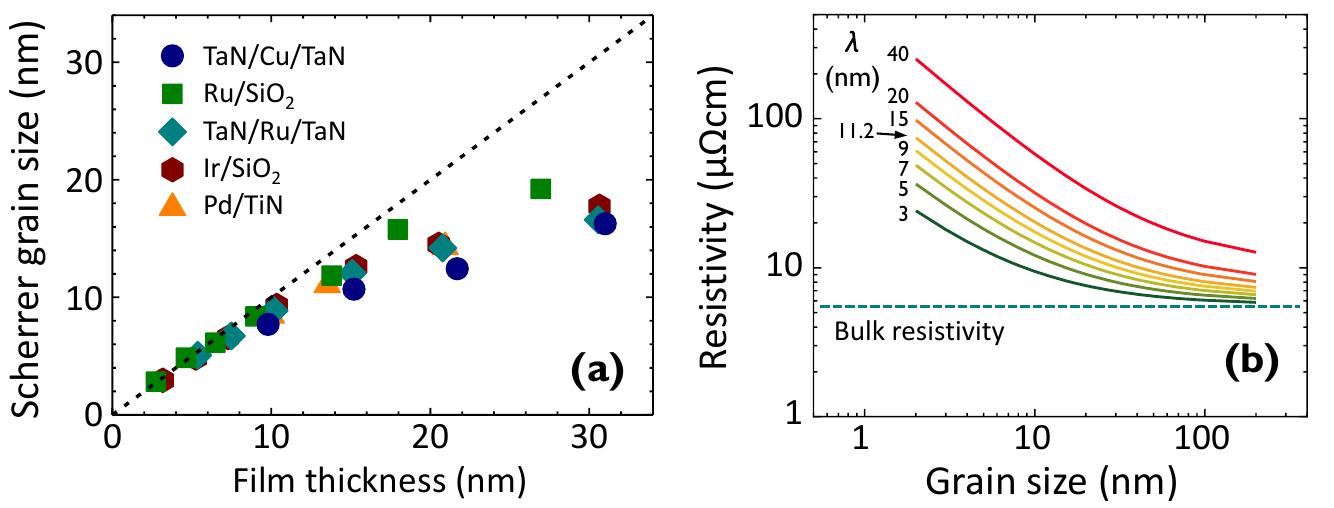}
\caption{\label{fig:MFP_MS_simul} (a) Experimental Scherrer grain size \textit{vs.}\ film thickness for various metals deposited by physical vapor deposition and annealed at 420\deg C. Reprinted from Ref.~\onlinecite{dutta_thickness_2017}. (b) Calculated thin film resistivity \textit{vs.}\ grain size using Eq.~\eqref{Eq:MS}. The parameters employed were: $\rho_0 = 5.3$ $\mu\Omega$cm, $R = 0.45$, $p = 0$ (diffuse surface scattering), and film thickness $h = 20$ nm.  These parameters, aside from the mean free path $\lambda$, are representative of Mo. For bulk Mo, $\lambda = 11.2$ nm.\cite{gall_electron_2016} Generally, a shorter mean free path $\lambda$ results in a weaker dependence of the thin film resistivity on grain size.}
\end{figure}

Equation \eqref{Eq:MS} contains five independent material parameters: bulk single-crystal resistivity, electron mean free path, surface/interface scattering specularity, average grain size, and the grain boundary reflection coefficient. In practice, the grain size often depends strongly on film thickness, introducing a secondary indirect source of thickness dependence of thin film resistivity, in addition to surface scattering effects. The relationship between grain size and film thickness, $d (h)$, is highly sensitive to deposition parameters and is not an intrinsic property of the material. Although a linear dependence has historically been assumed in some cases, experimental data typically do not support this assumption over a wide range of thicknesses (see Fig.~\ref{fig:MFP_MS_simul}a). Post-deposition annealing can further influence grain size, promoting grain growth or recrystallization.\cite{doherty_current_1997,humphreys_recrystallization_2004}

As a consequence, it is not possible to quantitatively predict the resistivity of metallic nanostructures based solely on bulk properties. As previously mentioned, the relationship $d (h)$ is strongly influenced by deposition process parameters and thermal budget, and no predictive models currently exist for grain structures and grain sizes beyond general trends.\cite{thompson_structure_2000,petrov_microstructural_2003} Given that experimental data indicate that grain boundary scattering can dominate thin film resistivity,\cite{dutta_thickness_2017} the evaluation of metals should consider the potential for achieving large-grain films (or nanowires). However, this cannot be reliably predicted by \textit{ab initio} calculations, requiring experimental studies on grain size and annealing behavior to complement transport and reliability metrics for a more accurate assessment of alternative metals.

A second limitation is the difficulty in calculating or predicting realistic values for the parameters $R$ and $p$. The parameter $R$ represents the electron reflection probability at a grain boundary, which generally depends on the relative orientation of grains on either side of the boundary and the atomic configuration of the grain boundary itself. Both theoretical calculations\cite{zhou_first-principles_2022,feldman_simulation_2010,zhou_ab_2010,cesar_calculated_2014} and experimental studies\cite{kim_large_2010,kim_structural_2011} consistently show significant variations in $R$ between small-angle or coincidence grain boundaries (with low $R$) and large-angle random grain boundaries (with higher $R$). In polycrystalline films, the relevant value of $R$ is an effective average across all grain boundary configurations, making it strongly dependent on the film's microstructure. While some trends suggest that larger cohesive energies lead to higher $R$ values,\cite{zhu_electron_2010} $R$ should not be treated as an intrinsic property of the metal alone.

Similarly, the surface scattering specularity, $p$, is influenced not only by the metal itself but also by the properties of the adjacent surfaces or interfaces. The cladding material\cite{rossnagel_alteration_2004,tay_electrical_2005,feldman_calculation_2009,ke_resistivity_2009} and surface or interface roughness are expected to play critical roles. While theoretical\cite{feldman_dependence_2008,timoshevskii_influence_2008,ke_resistivity_2009,rickman_resistivity_2012,moors_resistivity_2014,moors_modeling_2015,zhou_electrical_2018} and experimental\cite{marom_effect_2006,feldman_dependence_2008} studies have explored surface scattering as a function of surface roughness, quantitative agreement between theory and experiment remains elusive. Hence, predictive calculations for screening purposes are not feasible. Furthermore, because surface scattering depends on both the cladding material and surface roughness, it cannot be regarded as an inherent material property neither.

However, the previous Sec.~\ref{Sec:Semiclassical} demonstrates that the increase in resistivity, whether due to surface or grain boundary scattering, scales with the bulk mean free path of the charge carriers, $\lambda$. Therefore, the effects of high $R$, small grains, or diffuse interfaces can be mitigated by a short $\lambda$. This is illustrated in Fig.~\ref{fig:MFP_MS_simul}b, which shows the dependence of thin film resistivity on grain size, calculated using Eq.~\eqref{Eq:MS}, for $\rho_0 = 5.3$ $\mu\Omega$cm, $R = 0.45$, $p = 0$ (diffuse surface scattering), and a fixed film thickness of 20 nm as an example. The results indicate that a shorter mean free path, $\lambda$, leads to a weaker dependence of resistivity on grain size, thereby reducing the impact of grain boundary scattering. A similar trend can be observed for the dependence of thin film resistivity on thickness in the presence of diffuse surface scattering.

This insight has driven the search for metals with short $\lambda$ as potential alternatives to Cu. At room temperature, the mean free path of Cu is as high as 40 nm,\cite{gall_electron_2016} which is large compared to typical state-of-the-art interconnect dimensions and grain sizes. Therefore, metals with much shorter $\lambda$ promise to be less sensitive to interconnect scaling. An \textit{ab initio} methodology for screening short-$\lambda$ metals will be introduced in Sec.~\ref{Subsec:Rholambda}. The complete alternative metal screening process and its current status will be detailed in Sec.~\ref{Sec:Downselection}.

\subsubsection{Influence of resistivity anisotropy on thin film resistivity scaling\label{Sec:AnisoModel}}

The thin film resistivity models discussed above assume a spherical isotropic Fermi surface, effectively treating the metal as a free electron gas with an effective mass. Consequently, both bulk and thin film resistivities are independent of the crystallographic direction of the current, enabling the derivation of the Fuchs--Sondheimer and the Mayadas--Shatzkes equations in Eqs.~\eqref{eq:fuchs} and  \eqref{Eq:MS}, respectively. Due to the inherent complexity of the problem, no thin film resistivity model currently exists that accounts for the detailed band structure of the metal while incorporating both surface and grain boundary scattering. While the Mayadas--Shatzkes model in Eq.~\eqref{Eq:MS} has been able to successfully describe experimental measurements,\cite{learn_resistivity_1985,zhang_influence_2006, zhang_influence_2006b, camacho_surface_2006, sun_dominant_2009, sun_surface_2010, chawla_electron_2011, barmak_surface_2014, dutta_thickness_2017, founta_properties_2022} the lack of a quantitative understanding of key model parameters, such as $R$ and $p$, limits the ability to accurately evaluate the validity of the spherical Fermi surface approximation across different metals.

From a macroscopic perspective, the resistivity of cubic crystal systems is isotropic, making the approximation of a spherical Fermi surface potentially valid for such metals. However, in less symmetric structures (\textit{e.g.}, hexagonal, tetragonal, orthorhombic, monoclinic, trigonal), the resistivity becomes anisotropic in the bulk metal. For example, hexagonal Ru exhibits lower resistivity along the hexagonal axis compared to the two perpendicular (in-plane) directions.\cite{volkenshtejn_electric_1978,savitskii_research_1979} To address this, a semiclassical model based on the Mayadas--Shatzkes framework was developed for ellipsoidal Fermi surfaces.\cite{de_clercq_resistivity_2018} This model can describe metals with hexagonal, tetragonal, or orthorhombic crystal symmetries. While the mathematical details are beyond the scope of this tutorial, the model reveals a significant influence of Fermi surface anisotropy on surface scattering, without affecting grain boundary scattering.

\begin{figure}
\includegraphics[width=8cm]{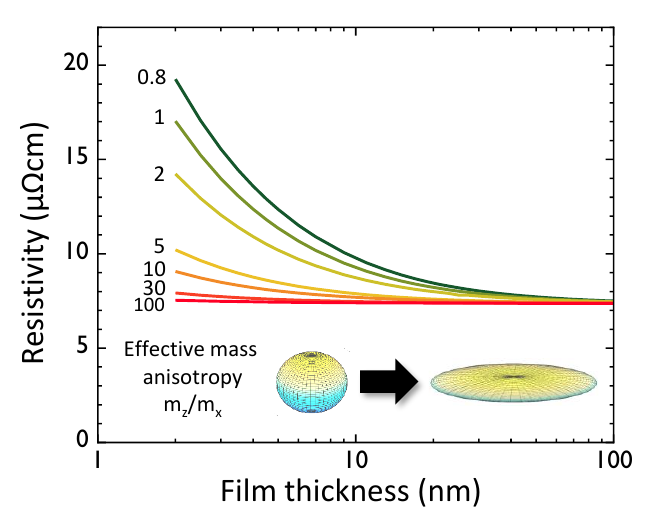} \caption{\label{fig:Aniso_simul} Calculated thin film resistivity \textit{vs.}\ film thickness for various conduction band effective mass anisotropies, considering surface scattering effects. The model details are provided in Ref.~\onlinecite{de_clercq_resistivity_2018}. The parameters employed were: $\rho_0 = 7.4$ $\mu\Omega$cm, $\lambda = 6.8$ nm, $R = 0$ (no grain boundary scattering), and $p = 0$ (diffuse surface scattering). These parameters are representative of Ru. For bulk Ru, the intrinsic effective mass anisotropy is 0.8. The results demonstrate that an oblate anisotropy, characterized by low in-plane resistivity (low in-plane effective mass) and high out-of-plane resistivity (high out-of-plane effective mass), reduces surface scattering and consequently yields a weaker dependence of the thin film resistivity on thickness.} 
\end{figure}

This behavior is illustrated in Fig.~\ref{fig:Aniso_simul}, which shows the thin film resistivity as a function of film thickness, with varying degrees of Fermi surface anisotropy, based on the model from Ref.~\onlinecite{de_clercq_resistivity_2018}. In these simulations, grain boundary scattering was neglected ($R = 0$), allowing surface scattering to dominate. The results demonstrate that in metals with low in-plane resistivity (corresponding to a small in-plane effective mass) and high out-of-plane resistivity (large out-of-plane effective mass), surface scattering is progressively suppressed, leading to a reduction in the thickness dependence of the thin film resistivity due to surface scattering.

In principle, this anisotropy effect could be exploited to reduce the resistivity of nanowires, and, therefore, the use of two-dimensional and one-dimensional metals (with a single low-resistivity crystallographic direction) in interconnect applications has been proposed.\cite{kumar_ultralow_2022} However, two important considerations must be noted. First, since Matthiessen's rule does not generally apply, the presence of significant grain boundary scattering—--which is unaffected by reduced dimensionality—--can suppress the benefits of surface scattering reduction in two- or one-dimensional metals, particularly in small-grain microstructures. Second, the application of one-dimensional metals in interconnects would necessitate single-crystal materials to ensure that the current is always aligned with the low-resistivity crystallographic direction. Currently, there is no feasible integration route for incorporating single-crystal lines in commercial interconnects, limiting the practical application of such materials to fundamental material science at this stage.

\subsection{\label{Sec:Resistivity_Intermet}Point defects, disorder, and alloy scattering in compound metals}

Before introducing first-principles screening methodologies for identifying and selecting promising metals for interconnect applications based on the mean free path of charge carriers, $\lambda$, it is essential to briefly discuss additional sources of scattering that are particularly relevant for compound metals. In crystalline materials, any deviation from periodicity can result in electron scattering.\cite{ashcroft_solid_1976} In elemental metals, such deviations include vacancy or vacancy cluster defects as well as impurities. In high-quality polycrystalline thin films of relevant metals, vacancies and impurities primarily influence the resistivity at cryogenic temperatures, but their effects are generally negligible at room temperature, where scattering by  phonons, grain boundaries, and surfaces dominates.

\begin{figure}
\includegraphics[width=8cm]{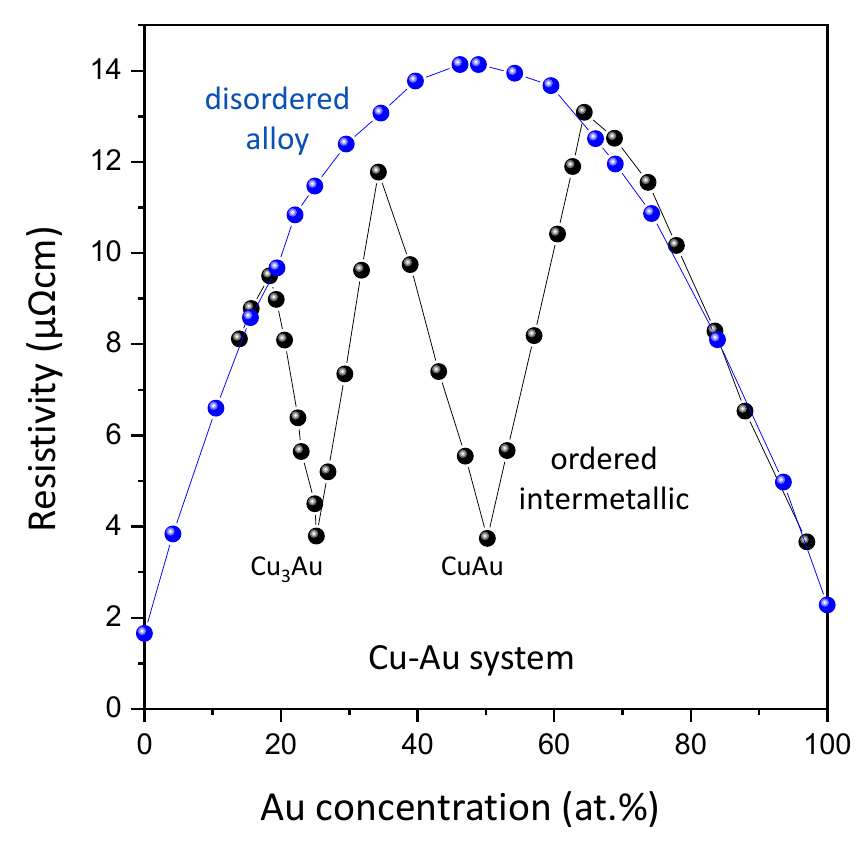} \caption{\label{fig:alloy_intermetallic} Resistivity of Cu$_{x}$Au$_{1-x}$ intermetallic compounds as a function of Au mole fraction. Randomly disordered alloys exhibit much higher resistivities due to alloy scattering, with a resistivity maximum at 50\%{} Au. In contrast, ordered Cu$_3$Au and CuAu intermetallics display significantly lower resistivities near their stoichiometric compositions.\cite{Johansson_CuAu_1936,terada_thermal_2002}} 
\end{figure}

The situation can however be markedly different for compound metals. Alloys are inherently disordered materials, characterized by the random distribution of different atoms on lattice sites. As a result, the crystal lacks periodicity, leading to increased resistivity due to alloy scattering. This effect is illustrated in Fig.~\ref{fig:alloy_intermetallic} for the seminal Cu--Au system.\cite{Johansson_CuAu_1936} In disordered Cu$_x$Au$_{1-x}$ alloys, the resistivity is considerably higher than in the pure elemental metals Cu and Au, reaching a maximum at 50\%{} Au content, where disorder is greatest. By contrast, the system also forms two ordered intermetallic phases, Cu$_3$Au and CuAu, near their respective stoichiometries. In these ordered phases, the resistivity shows a sharp minimum, significantly lower than that of the disordered alloys.

These observations highlight the necessity of minimizing alloy scattering to achieve resistivities relevant for interconnect applications in compound metals. As a result, in addition to elemental metals, for which such issues do not arise, only ordered intermetallics are of potential interest in alternative metal screening efforts. This challenge is further compounded by the difficulty of accurately measuring and optimizing intermetallic ordering in thin films. A more detailed discussion and the current state of the art can be found in Secs.~\ref{Sec:binary} and \ref{Sec:ternary}.

\subsection{\textit{Ab initio} screening of alternative metals\label{Subsec:Rholambda}}

The discussions in Secs.~\ref{Sec:Res_Mat_Dep} and \ref{Sec:AnisoModel} emphasize that accurately predicting the resistivity of thin films or nanowires is not feasible without detailed microstructural information, which is inherently dependent on deposition process conditions. Additionally, resistivity models such as the Mayadas--Shatzkes model in Eq.~\eqref{Eq:MS} rely on the assumption of a spherical Fermi surface (free electron gas), and metal-specific scaling is represented solely by a single mean free path value. While it is possible to compute the electron--phonon-limited resistivity of metals from first principles by considering detailed band structures,\cite{Ponce2020} such calculations remain computationally intensive, restricting feasible system sizes to only a few atoms per unit cell. The primary limitation is the computational cost of calculating the electron--phonon coupling and the corresponding relaxation times. Moreover, incorporating grain boundaries into this framework is highly challenging. To date, these factors constrain the development of a fully predictive downselection methodology capable of directly identifying the most promising metal candidates for interconnect applications.

Consequently, the metal selection problem must be approached in stages, as outlined in Fig.~\ref{fig:workflow}. Nevertheless, screening metals of potential interest remains feasible using a $\rho_0\times\lambda$ figure of merit, which can be computed using \textit{ab initio} methods with relatively low computational cost, as introduced below. The application of this methodology for screening elemental and compound metals will be discussed in Sec.~\ref{Sec:Downselection}.

\begin{figure}
\includegraphics[width=0.49\textwidth]{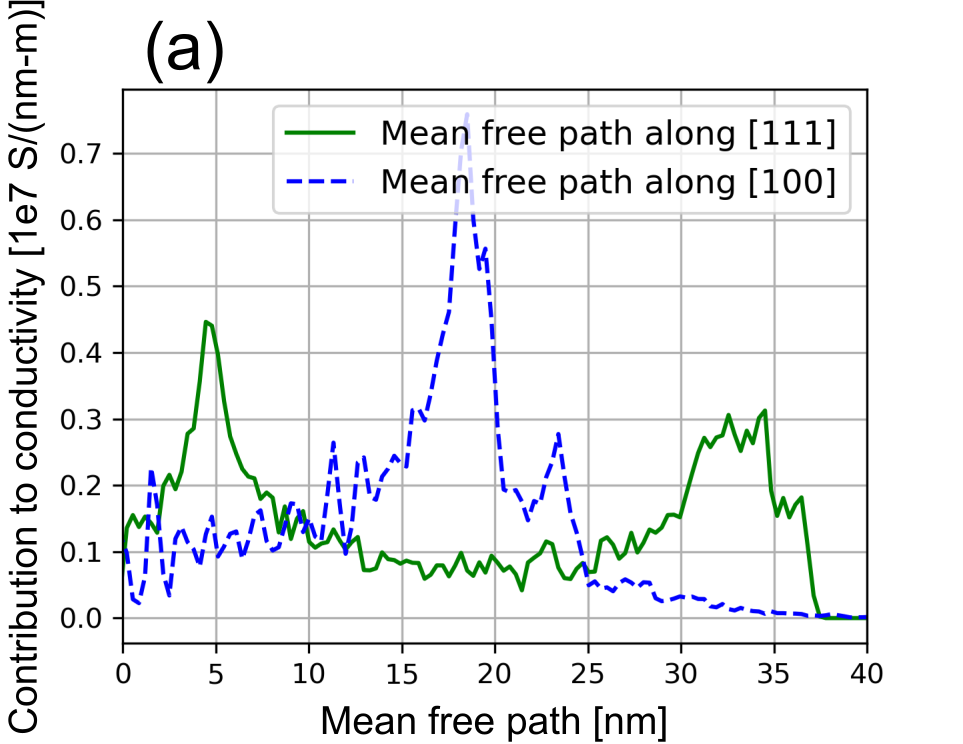}
\includegraphics[width=0.46\textwidth]{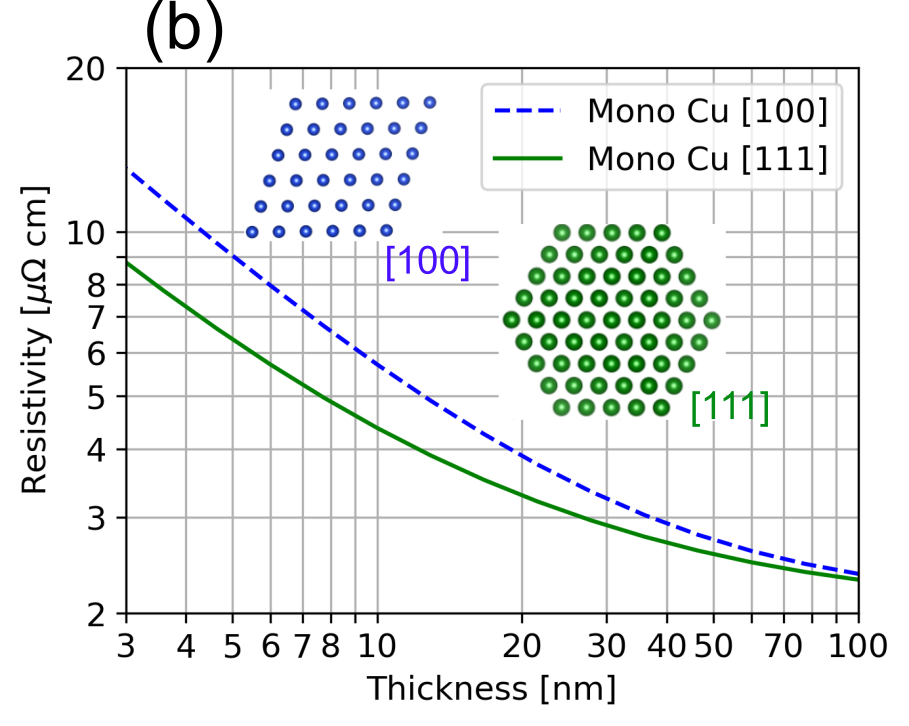} 
\caption{(a) Decomposition of the electrical conductivity for Cu as a function of the mean free path along the [111] and [100] crystallographic directions. The distinct profiles observed along these two directions can be attributed to the Fermi surface anisotropy. (b) \textit{Ab initio} calculations of the film thickness dependence of the resistivity for monocrystalline Cu with two different surface normal orientations considering surface scattering effects. Computational details are the same as those in Ref.~\onlinecite{VanTroeye2023}. 
\label{fig:meanfree}}
\end{figure}

In the Boltzmann transport framework, the conductivity tensor for a bulk metal film is expressed by Eq.~\eqref{eq:sigma} as 

\begin{equation}
\label{Eq:transport}
\underline{\underline{\sigma}} = \frac{-e^2}{N_\mathbf{k}\Omega}  \sum_{n\mathbf{k}}   \tau_{n\mathbf{k}} \frac{\partial f^0_{n\mathbf{k}}}{\partial \epsilon_{n\mathbf{k}}} \mathbf{v}_{n\mathbf{k}} \otimes \mathbf{v}_{n\mathbf{k}}.
\end{equation}

\noindent As noted above, the calculation of the conductivity tensor requires the knowledge of the the relaxation time $\tau_{n\mathbf{k}}$, which is costly to calculate by \textit{ab initio} methods. In general, $\tau_{n\mathbf{k}}$ of an electron depends on its wavevector $\mathbf{k}$ and band index $n$; therefore, metals typically show a broad distribution of mean free paths,\cite{chen_interdiffusion_2021,VanTroeye2023} as illustrated in Fig.~\ref{fig:meanfree}a for Cu along two crystallographic directions. 

Nevertheless, the complexity of the problem can be reduced by assuming an isotropic constant relaxation time $\tau$, independent of the electron wavevector $\mathbf{k}$. Under this approximation, Eq.~\eqref{Eq:transport} can be simplified by extracting $\tau$ from the summation and evaluating the expression only at the Fermi energy, \textit{i.e.}, by replacing the derivative of the Fermi--Dirac distribution by a $\delta$-function. This results in the transport tensor\cite{gall_electron_2016,moors_first-principles-based_2022}

\begin{equation}
	\label{eq:rho_tau}
	\underline{\underline{\frac{1}{\rho_0 \tau}}} = \frac{e^2}{(2 \pi)^3} \sum_n \left\langle \mathbf{v}^{(n)}(\mathbf{k}) \otimes \mathbf{v}^{(n)}(\mathbf{k}) \right\rangle_\mathrm{BZ},
\end{equation}

\noindent where BZ denotes that the summation is performed over the Brillouin zone. The approximation of the Fermi--Dirac distribution to a $\delta$-function has been found to be generally accurate, resulting in temperature-independent transport tensors.\cite{moors_first-principles-based_2022}

The key advantage of the constant relaxation time approximation is that the right-hand side of Eq.~\eqref{eq:rho_tau} depends only on the morphology of the Fermi surface, eliminating the need for detailed knowledge of electron--phonon interactions. As a result, the calculation of the $\underline{\underline{\rho_0 \tau}}$ transport tensor is computationally much less demanding, rendering it suitable for evaluating a wide range of metals.\cite{gall_electron_2016,moors_first-principles-based_2022}

Alternatively, an equivalent transport tensor can be formulated by assuming that the mean free path of the charge carriers $\lambda(\mathbf{k})=\mathbf{v}^{(n)}\times\tau^{(n)}(\mathbf{k})$ is isotropic and independent of $\mathbf{k}$. This approximation, the constant mean free path approximation, leads to the following transport tensor:\cite{gall_electron_2016,moors_first-principles-based_2022}

\begin{equation}
	\label{eq:rho_lambda}
	\underline{\underline{\frac{1}{\rho_0 \lambda}}} = \frac{e^2}{(2 \pi)^3} \sum_n \left\langle \frac{\mathbf{v}^{(n)}(\mathbf{k}) \otimes \mathbf{v}^{(n)}(\mathbf{k})}{|\mathbf{v}^{(n)}(\mathbf{k})|} \right\rangle_\mathrm{BZ}.
\end{equation}

\noindent It is worth noting that a $\underline{\underline{\rho_0 \lambda}}$ tensor can also be derived within the constant relaxation time approximation by dividing the $\underline{\underline{\rho_0 \tau}}$ transport tensor by the Fermi--Dirac weighted average velocity

\begin{equation}
v \equiv \frac{\sum_n \langle |\mathbf{v}^{(n)}(\mathbf{k})| \rangle_\textsc{bz}}{\sum_n \langle 1 \rangle_\textsc{bz}}.
\end{equation}

\noindent While the numerical values of the various transport tensors may exhibit discrepancies for a given metal due to the inherent approximations, screening methodologies based on these tensors generally yield consistent results. Consequently, both approaches can be used interchangeably for screening purposes.

One practical application of this approach involves the approximate determination of a single-valued mean free path $\lambda$ (or relaxation time $\tau$) of a metal when the bulk resistivity $\rho_0$ is known, for instance, from experimental data. It is noteworthy that for cubic systems, the $\underline{\underline{\rho_0 \lambda}}$ and $\underline{\underline{\rho_0 \tau}}$ tensors are diagonal and thus reduce to a single isotropic value.\cite{gall_electron_2016} Many semiclassical thin film transport models, such as the Mayadas--Shatzkes model in Eq.~\eqref{Eq:MS},\cite{mayadas_electrical-resistivity_1970} assume a simplified isotropic free electron gas and consequently neglect band structure effects, including their influence on the mean free path. For real metals with complex band structures, it may be possible to replace the exact $\mathbf{k}$-dependent mean free path by an effective mean free path. However, rigorous calculations of the electron--phonon coupling are required for accurate determination of this effective value. Therefore, the $\lambda$ value extracted from the $\underline{\underline{\rho_0 \lambda}}$ tensor (divided by the bulk resistivity $\rho_0$) has been utilized as an effective mean free path in thin film transport models. Given the experimental challenges associated with directly measuring the mean free path,\cite{krewer_thickness-dependent_2020} the accuracy of this approximation remains uncertain. Nevertheless, employing $\underline{\underline{\rho_0 \lambda}}$-derived values for the mean free path has generally led to satisfactory agreement between semiclassical transport models and experimental observations.\cite{learn_resistivity_1985,zhang_influence_2006, zhang_influence_2006b, camacho_surface_2006, sun_dominant_2009, sun_surface_2010, chawla_electron_2011, barmak_surface_2014, dutta_thickness_2017, founta_properties_2022} 

Secondly, the  $\underline{\underline{{\rho_0 \lambda}}}$ tensor can serve as a figure of merit for a metal, indicating its potential for achieving low resistivity at nanoscale dimensions. Lower values of this tensor correspond to metals with greater scaling potential. Consequently, both the $\underline{\underline{{\rho_0 \lambda}}}$ and $\underline{\underline{\rho_0 \tau}}$ tensors have been extensively employed to identify promising alternative metals for nanoscale interconnect applications. A significant advantage of this approach lies in its substantially reduced computational cost compared to independently calculating the electron--phonon mean free path and bulk resistivity, enabling the screening of a wide range of materials. However, given the inherent approximations, the screening results should be interpreted with caution, as they may not accurately predict the thin film resistivity (see also Sec.~\ref{Sec:Res_Mat_Dep}). Nevertheless, this screening methodology has proven valuable, and the current state-of-the-art techniques utilizing this approach for alternative metal screening are discussed in Sec.~\ref{Sec:Downselection}.

\subsection{Resistivity scaling for thin films and nanowires in presence of surface scattering\label{Sec:TF_NW}}

While the methodology introduced in the previous section has been successfully applied to screening both elemental and compound metals, its accuracy in predicting thin film or nanowire resistivities is limited. In reality, the mean free path of charge carriers depends strongly on their wavevector $\mathbf{k}$. As depicted in Fig.~\ref{fig:meanfree}a for Cu, the mean free path varies significantly along different directions on the Fermi surface. Consequently, incorporating the complete anisotropic band structure into transport calculations and screening efforts is essential for achieving a more accurate understanding of resistivity at nanoscale dimensions.

Equation~\eqref{eq:rhoh} captures the rescaling of the bulk relaxation time, accounting for all pertinent scattering mechanisms. It should be noted that grain boundary scattering can be implicitly accounted for through an effective mean free path. Under the assumption of a spherical Fermi surface, the analytical Mayadas--Shatzkes model, as expressed in Eq.~\eqref{Eq:MS}, can be derived. Incorporating the full electronic band structure yields the following expression:

\begin{equation}
    \frac{\tau_{n\mathbf{k}}}{\tau^0_{n\mathbf{k}}} =1+\frac{|\boldsymbol{\lambda}_{n\mathbf{k}}.\hat{u}|}{h} \frac{1-p}{1-p \exp{\left(\frac{-h}{|\boldsymbol{\lambda}_{n\mathbf{k}}.\hat{u}|}\right)}}\left[1-\exp{\left\{\frac{-h}{|\boldsymbol{\lambda}_{n\mathbf{k}}.\hat{u}|}\right\}}\right]~\label{eq:modifiedtau2}.
\end{equation}

\noindent Here, $\tau^0_{n\mathbf{k}}$ represents the relaxation time, accounting for both phonon and grain boundary scattering. 

When grain boundary scattering is neglected (\textit{i.e.}, considering only phonon and surface scattering), the thin film resistivity can be predicted \textit{ab initio}. The calculated relationship between thin film resistivity and film thickness for Cu is shown in Fig.~\ref{fig:meanfree}b. It is important to note that Eq.~\eqref{eq:modifiedtau2} indicates that the resistivity of thin films (or nanowires) depends not only on the transport direction but also on the orientation of the surface normal (growth orientation). This holds true even for metals with isotropic bulk resistivity, such as cubic metals like Cu, and can be explained by the reduced symmetry arising from dimensional confinement. 

For single-crystal films with negligible grain boundary scattering, the results are, in principle, exact, apart from the usual approximations inherent to density functional theory (DFT). While grain boundary scattering in textured films could theoretically be incorporated via a grain-size-dependent mean free path, its accurate treatment relies heavily on the knowledge of the detailed microstructure and remains challenging for current \textit{ab initio} techniques.

Beyond transport calculations, the proposed model can also be extended to derive a figure of merit for thin films or nanowires, analogous to the $\underline{\underline{{\rho_0 \lambda}}}$ tensor in Eq.~\eqref{eq:rho_lambda}. For a thin film with surface normal $\mathbf{\hat{u}}$, we find:~\cite{VanTroeye2023}

\begin{equation}
    \left. \frac{1}{\rho\lambda(\mathbf{\hat{u})}}\right|_\mathrm{film} =  -\frac{n_se^2}{N_\mathbf{k}\Omega}\sum_{|\mathbf{\hat{v}}_{n\mathbf{k}}\cdot\mathbf{\hat{u}}|>\theta} \frac{\partial f_{n,\mathbf{k}}}{\partial \varepsilon} \frac{|\mathbf{\mathbf{v}}_{n,\mathbf{k}}\cdot\mathbf{\hat{n}}|^2}{2|\mathbf{v}_{n\mathbf{k}}||\mathbf{\hat{v}}_{n\mathbf{k}}\cdot\mathbf{\hat{u}}|} \label{eq:rholambdau_tf}.
\end{equation}

\noindent To prevent divergence, an angular cutoff $\theta$ was introduced. The unit vector $\mathbf{\hat{n}}$ represents the in-plane transport direction. 

The expression for a nanowire is given by\cite{VanTroeye2023}

\begin{equation}
    \left. \frac{1}{\rho\lambda(\mathbf{\hat{u}},\mathbf{\hat{s}})} \right|_\mathrm{wire} =  -\frac{n_se^2}{N_\mathbf{k}\Omega}\sum_{|\mathbf{\hat{v}}_{n\mathbf{k}}\cdot\mathbf{\hat{u}}|>\theta,|\mathbf{\hat{v}}_{n\mathbf{k}}\cdot\mathbf{\hat{s}}|>\theta} \frac{\partial f_{n,\mathbf{k}}}{\partial \varepsilon} |\mathbf{v}_{n,\mathbf{k}}\cdot\mathbf{\hat{n}}|^2 \tau^{nw}_{n\mathbf{k}}\label{eq:rholambdau_bvt}, 
\end{equation}

\noindent with $\tau^{nw}_{n\mathbf{k}}$ given by the expression\cite{VanTroeye2023}

\begin{equation}
  \begin{cases}
    \tau^{nw}_{n\mathbf{k}} \approx \frac{h}{2|\mathbf{v}_{n\mathbf{k}}\cdot\mathbf{\hat{u}}|} \left(1-\frac{h}{w}\frac{|\mathbf{v}_{n\mathbf{k}}\cdot\mathbf{\hat{u}}|}{|\mathbf{v}_{n\mathbf{k}}\cdot\mathbf{\hat{s}}|}\right) + O(h^3)
    \text{ if }w |\mathbf{v}_{n\mathbf{k}}\cdot\mathbf{\hat{u}}| < h |\mathbf{v}_{n\mathbf{k}}\cdot\mathbf{\hat{s}}| \\
    \tau^{nw}_{n\mathbf{k}} \approx \frac{w}{2|\mathbf{v}_{n\mathbf{k}}\cdot\mathbf{\hat{u}}|} \left(1-\frac{w}{h}\frac{|\mathbf{v}_{n\mathbf{k}}\cdot\mathbf{\hat{s}}|}{|\mathbf{v}_{n\mathbf{k}}\cdot\mathbf{\hat{u}}|}\right) + O(h^3)
    \text{ if }w |\mathbf{v}_{n\mathbf{k}}\cdot\mathbf{\hat{u}}| > h |\mathbf{v}_{n\mathbf{k}}\cdot\mathbf{\hat{s}}| \\
    \tau^{nw}_{n\mathbf{k}} \approx \frac{h}{3|\mathbf{v}_{n\mathbf{k}}\cdot\mathbf{\hat{u}}|} + O(h^4) \text{ if }w |\mathbf{v}_{n\mathbf{k}}\cdot\mathbf{\hat{u}}| = h |\mathbf{v}_{n\mathbf{k}}\cdot\mathbf{\hat{s}}|. \\
  \end{cases} \label{eq:rholambda_2D}
\end{equation}

\noindent Here, $\mathbf{\hat{s}}$ and $\mathbf{\hat{u}}$ denote the two confinement directions corresponding to the width $w$ and thickness $h$, respectively. An alternative formulation to the thin film and nanowire figures of merit has been proposed in Ref.~\onlinecite{kumar_ultralow_2022} and leads to comparable results in terms of metal benchmarking.

This figure of merit accounts for anisotropic transport effects, such as the suppression of surface scattering discussed in Sec.~\ref{Sec:AnisoModel}, making it particularly suitable for single-crystal wires, in which transport occurs in well-defined crystallographic directions. For certain crystalline orientations, this can result in highly favorable figures of merit for specific materials, in particular one-dimensional metals.\cite{kumar_ultralow_2022} However, it is important to note that in random polycrystalline films and nanowires, surface scattering is averaged over all grain orientations. Additionally, in films or nanowires where resistivity is predominantly governed by grain boundary scattering (\textit{e.g.}, in small-grain polycrystalline films), the suppression of surface scattering is significantly reduced (see Sec.~\ref{Sec:AnisoModel}). Under these conditions, benchmarking metals using Eq.~\eqref{eq:rho_lambda} remains the most appropriate approach.

\section{Interconnect reliability\label{Sec:Reliability}}

A second critical aspect in selecting new metallization schemes for interconnects is the reliability of both the metal and the surrounding dielectric materials. Interconnect failure can originate from degradation in either the metal or the dielectric. Analogous to line resistance, interconnect reliability tends to deteriorate as wires and vias are miniaturized. Initially, the limited electromigration resistance of Al led to the adoption of Cu as the primary interconnect metallization more than two decades ago.\cite{edelstein_full_1997} However, the reliability of Cu metallization is now facing increasing challenges. As elaborated in the subsequent section, barrier and liner layers are essential for maintaining reliability but must be scaled alongside interconnect dimensions to leave sufficient room for the Cu conductor. Due to the limited scalability of these layers, they eventually occupy a substantial fraction of the interconnect volume while contributing minimally to overall conductance. As demonstrated in Sec.~\ref{Sec:Interconnect_Model} using calibrated line resistance models, barrier- and liner-less metallization is essential for realizing low-resistance interconnects, outperforming current Cu-based metallization schemes. In the following sections, we will delve into the fundamentals of both metal and dielectric reliability. The current state of the art regarding the reliability of specific elemental and binary alternative metals will be explored in Sec.~\ref{Sec:Downselection}.

\subsection{Dielectric breakdown and the need for barrier layers}

Time-dependent dielectric breakdown (TDDB) is a physical degradation process in which a dielectric material, subjected to a constant electric field below its intrinsic breakdown strength, progressively deteriorates and ultimately fails over time. This failure can be attributed to the formation of conductive paths (filaments) within the dielectric, which short-circuit adjacent metallic electrodes.\cite{mcpherson_2012} Rapid failure may result either from gradual intrinsic damage to the dielectric material (\textit{e.g.}, vacancy formation) or from the drift of metal from nearby electrodes, such as interconnect lines or vias. In the latter scenario, metal ions drift through the dielectric under the influence of the applied electric field, contaminating the dielectric, increasing leakage currents, and ultimately causing dielectric breakdown and shorting between interconnect lines.\cite{wu_electrical_2014,croes_interconnect_2018}

The time-dependent dielectric breakdown behavior highly depends on the materials used, both metals and dielectrics. For metals, there exists a thermodynamic barrier that determines the detachment of ionized metal atoms and their subsequent drift or diffusion into the low-$\kappa$ dielectric. This barrier generally scales with the metal’s cohesive energy. Consequently, refractory metals, which exhibit significantly higher cohesive energies compared to Cu, present a higher barrier, thereby suppressing the drift and diffusion of metal ions into the dielectric and resulting in a substantially longer time-dependent dielectric breakdown lifetime. It is important to note that the drift and diffusion of metal ions \textit{within} the dielectric are less influenced by the cohesive energy of the metal and are instead primarily dependent on the binding energy of metal impurities in the dielectric. Nevertheless, for metals relevant to interconnects, the dominant thermodynamic barrier arises from the detachment energy. Therefore, metals with high cohesive energy (\textit{i.e.}, refractory metals) are expected to exhibit significantly improved time-dependent dielectric breakdown lifetimes.

Moreover, the selection of dielectrics also influences the time-dependent dielectric breakdown performance of interconnects. Advanced interconnects utilize low-$\kappa$ materials to minimize interconnect capacitance.\cite{ho_low_2003,shamiryan_low-k_2004,volksen_low_2010,grill_progress_2014} The low dielectric permittivity $\kappa$ of these dielectrics is typically associated with a reduced dipole density, either intrinsically due to their composition or by decreasing physical density, such as through the introduction of porosity. This can have a pronounced impact on time-dependent dielectric breakdown behavior, as it is well-established that Cu readily detaches and diffuses in such dielectrics, leading to rapid dielectric breakdown. As a result, Cu interconnects require the use of refractory barrier layers between the dielectric and the Cu metallization to prevent Cu detachment and drift into the dielectric. In general, amorphous barrier materials are preferred over polycrystalline counterparts, as grain boundaries can act as diffusion pathways. Currently, TaN is employed as the primary barrier material, with the potential for being scaled down to thickness around 1 to 1.5 nm without compromising functionality.\cite{witt_testing_2018,Pedreira_2019_TaN} Ongoing research has explored alternative barrier materials, such as Zn-doped Ru,\cite{joi_doped_2022} though their integration poses challenges and is not expected to yield significant improvements over TaN barriers.

\begin{figure}[b]
\includegraphics[width=14cm]{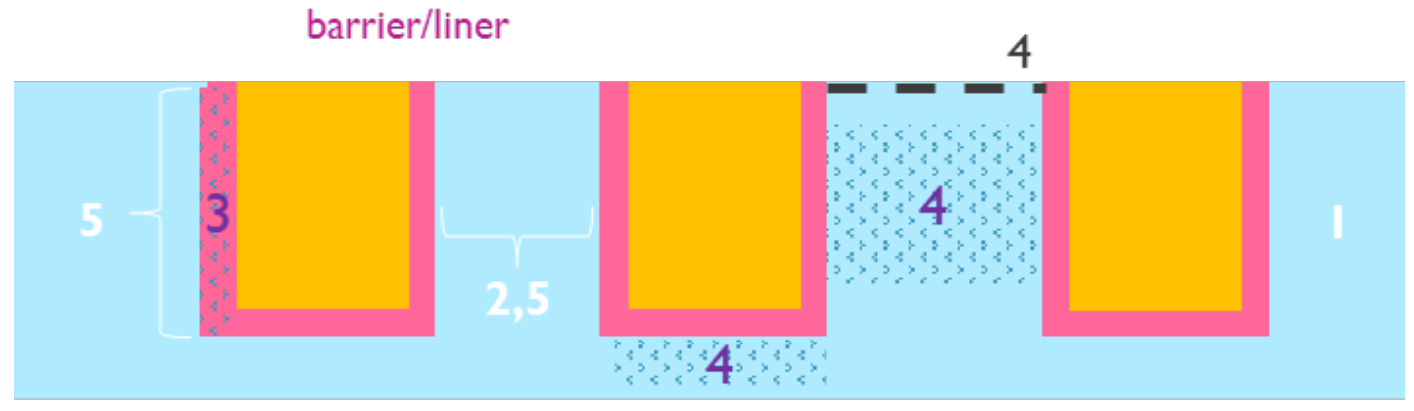} 
\caption{\label{Fig:TDDB-fig1} Illustration of factors influencing time-dependent dielectric breakdown and interconnect lifetime: (1) intrinsic dielectric properties, breakdown location, and mechanism; (2) insulating dielectric thickness; (3) barrier properties and conductor metal detachment; (4) dielectric damage induced by integration processes, such as barrier deposition, chemical-mechanical polishing, etching, or moisture absorption; (5) line variability, including line-edge roughness, trench height, or via misalignment.}
\end{figure}

Beyond intrinsic material properties, process limitations and dimensional scaling can further compromise time-dependent dielectric breakdown behavior. Factors such as narrow gaps, line-edge roughness, plasma-induced damage, and misalignment (see Fig.~\ref{Fig:TDDB-fig1}) can all contribute to dielectric degradation. Therefore, selecting a suitable alternative metal for advanced interconnects should not only be based on its intrinsic properties (such as the metal detachment barrier) but also on its compatibility with integration processes, as the introduction of new materials can exacerbate dielectric breakdown. It is thus crucial to experimentally evaluate time-dependent dielectric breakdown performance using appropriate test structures, such as planar capacitors (PCAPS)\cite{zhao_test_struc_2009} or sidewall capacitors (SWCAPS),\cite{lin_demonstration_2014} to facilitate the downselection of promising alternative metals.

The primary driving forces behind time-dependent dielectric breakdown are the applied electric field and the temperature of the interconnect. In practice, time-dependent dielectric breakdown constrains the maximum electric field that can be safely applied between adjacent lines, making it a critical consideration in circuit design and layout. For commercial circuits, lifetime requirements are typically set at 10 years for temperatures up to 135\deg{}C. Direct measurement of such long lifetimes is impractical, so the maximum electric field for reliable operation is determined through extrapolation from accelerated tests conducted at elevated temperatures and electric fields. There is ongoing debate in the literature regarding the most appropriate model for time-dependent dielectric breakdown lifetime extrapolation, with several proposed models:\cite{croes_low_2013}

\begin{align}
E\textrm{-model:}\qquad & t_\textrm{50\%}\propto\exp\left( -\gamma E\right)\\
\sqrt{E}\textrm{-model:}\qquad & t_\textrm{50\%}\propto\exp \left( -\gamma\sqrt{E}\right)\\
\textrm{Power-law model:}\qquad & t_\textrm{50\%}\propto E^{-m}\\
\textrm{Impact damage model:}\qquad & t_\textrm{50\%}\propto\frac{\exp\left( -\gamma\sqrt{E}+\frac{\alpha}{E}\right)}{E}\\
1/E\textrm{-model:}\qquad & t_\textrm{50\%}\propto\exp\left(\frac{\gamma}{E}\right)
\end{align}

\begin{figure}[tb]
\includegraphics[width=7cm]{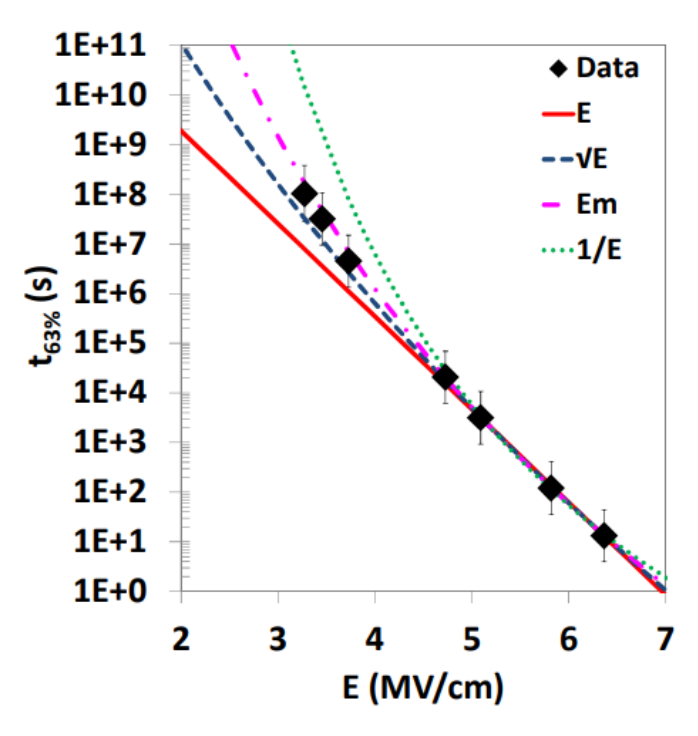} 
\caption{\label{Fig:TDDB-fig6} Comparison of available models for time-dependent dielectric breakdown lifetime extrapolation versus measured data for Cu. Reprinted with permission from Ref.~\onlinecite{croes_low_2013}.}
\end{figure}

\noindent Here, $t_\textrm{50\%}$ represents the time elapsed before a line fails with a probability of 50\%{}. The impact damage model, also known as the lucky electron model, is widely regarded to most accurately describe the underlying physical mechanisms of time-dependent dielectric breakdown.\cite{lloyd_simple_2005} However, some researchers prefer the power-law model for fitting time-dependent dielectric breakdown data, as it offers reliable predictions with a limited number of fitting parameters.\cite{croes_low_2013, jeong_low_2015, chery_identification_2013} In contrast, studies on damascene structures\cite{croes_e-_2010,liniger_low-field_2014} have found that both the $E$- and $\sqrt{E}$-models tend to be overly conservative in fitting low-field time-dependent dielectric breakdown data. The behavior of these models, along with their comparison to experimental data for Cu, is depicted in Fig.~\ref{Fig:TDDB-fig6}. It is important to note that additional area scaling and extrapolation to low failure percentiles are necessary to determine failure rates and define operating condition limits for industry-relevant interconnects.\cite{roussel_new_2018}

\begin{figure}[tb]
\includegraphics[width=0.9\textwidth]{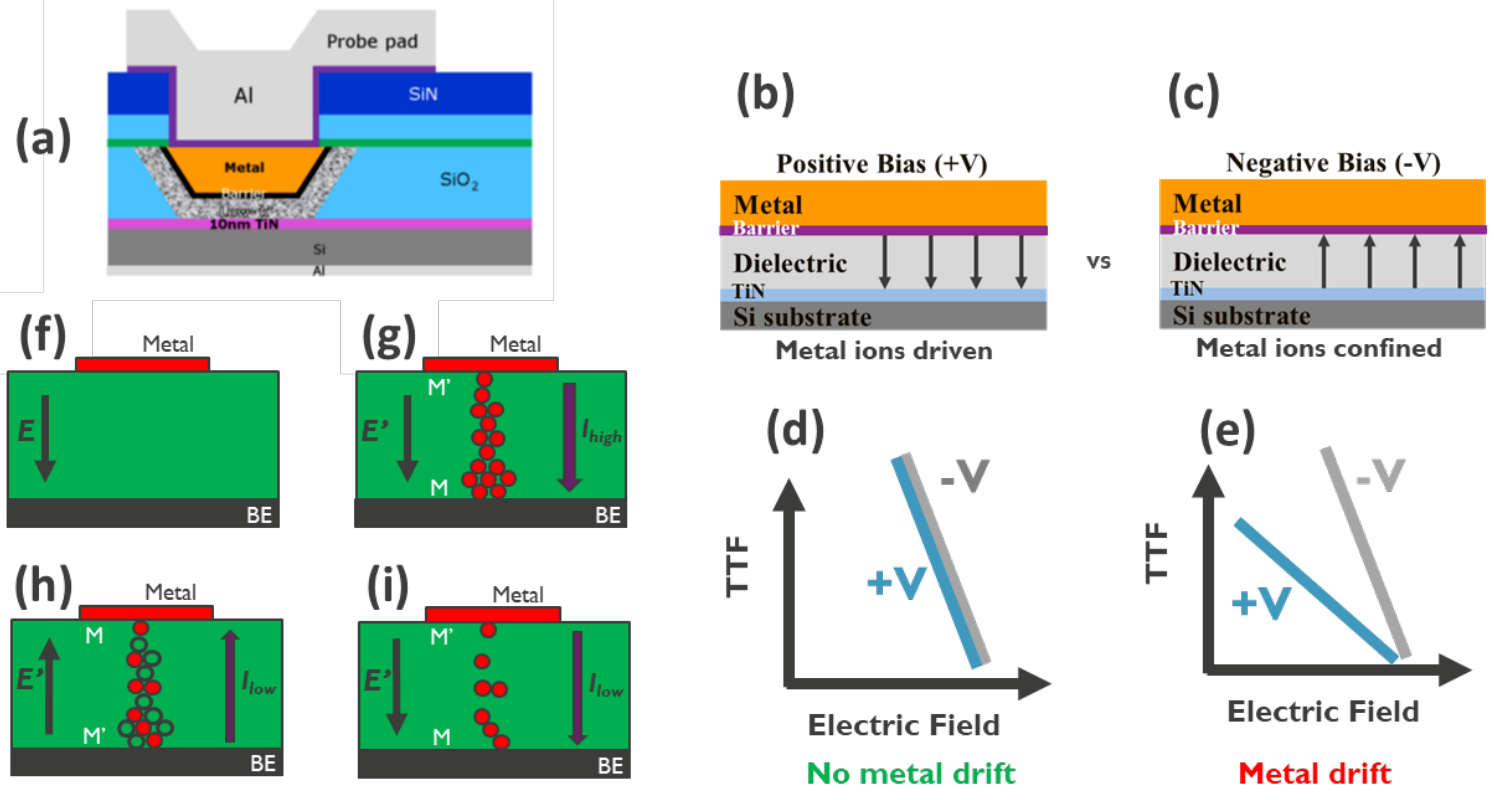} 
\caption{\label{Fig:TDDB_testing} (a) Device structure of a planar capacitor (PCAP) used for time-dependent dielectric breakdown lifetime measurements. (b) and (c) Schematics of metal drift processes under constant voltage stress. For weak metal top electrodes, metal ions can migrate during positive voltage, while metal drift does not occur under negative bias. (d) and (e) Time-to-failure (TTF) behavior as a function of applied bias voltage $\pm V$. (f) to (i) Illustration of metal filament formation and dissolution during triangular voltage sweeps with applied electric fields $E$ as indicated.}
\end{figure}

Accelerated time-dependent dielectric breakdown testing can also provide insights into the underlying failure mechanisms, such as whether breakdown occurs via dielectric failure or through the formation of metal filaments. Typical bias-temperature stress experiments employ PCAPs or SWCAPs at elevated temperatures, using methods such as triangular voltage sweeps  or the application of constant voltage (see Fig.~\ref{Fig:TDDB_testing}). For capacitors with different electrodes---one consisting of the metal of interest and the other of a refractory metal---metal drift can be identified and distinguished from intrinsic breakdown by comparing results under positive and negative bias. For instance, under positive voltage stress applied to the weak top electrode in a PCAP (Fig.~\ref{Fig:TDDB_testing}a), and in the absence of an effective diffusion barrier, metal ions will drift into the dielectric (Fig.~\ref{Fig:TDDB_testing}b). In contrast, no metal drift is observed under negative voltage stress (Fig.~\ref{Fig:TDDB_testing}c). This allows for the differentiation between metal drift and intrinsic dielectric breakdown mechanisms, as the latter are not dependent on bias polarity (Figs.~\ref{Fig:TDDB_testing}d and e).

Alternatively, during triangular voltage sweeps, metal drift can be detected when the leakage current increases during the initial sweep but disappears in subsequent sweeps, indicating that metal ions have migrated through the dielectric (Figs.~\ref{Fig:TDDB_testing}f to i). Time-dependent dielectric breakdown measurements conducted at various temperatures and voltages enable the study of conditions associated with intrinsic breakdown, metal filament growth, and filament formation.\cite{wu_insights_2018} Given that, these mechanisms are influenced by the metal, the dielectric material, and their respective thicknesses, it is crucial to continuously refine testing methodologies for alternative metals and advanced interconnects.

From a dielectric breakdown perspective, the primary limitation in scaling Cu interconnects is the requirement for barrier layers to prevent metal detachment and drift into the surrounding dielectrics. As interconnect dimensions shrink, these barriers---having significantly higher resistance than Cu---occupy a substantial fraction of the interconnect volume, which ultimately impedes effective metal fill at ultrasmall dimensions. As discussed further in Sec.~\ref{Sec:Interconnect_Model}, this results particularly in a sharp increase in line resistance for interconnects with widths below 10 nm, necessitating the adoption of barrierless alternative metallization strategies.

\subsection{Electromigration}\label{Electromigration}

\begin{figure}[tb]
\includegraphics[width=8cm]{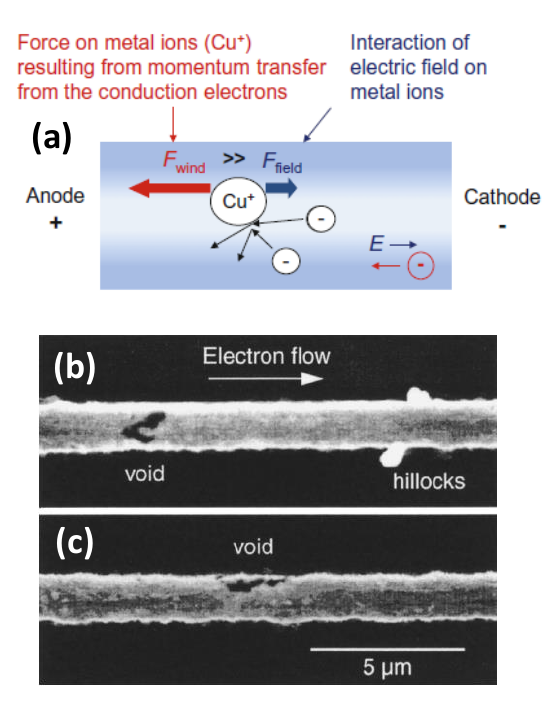} 
\caption{\label{Fig:Electrom_fig} (a) Schematic illustrating the electromigration driving force in a metal ($F_\mathrm{wind}$).\cite{Lienig_2005_VLSI} (b) and (c) Scanning electron micrographs of hillocks and voids induced by electromigration in a Cu line. Reprinted with permission from Ref.~\onlinecite{Ryu_1999_Trans}.}
\end{figure}

Electromigration (EM) is a well-documented main failure mechanism in integrated circuits. When an electric current flows through a conductor, the metal atoms are subjected to two opposing forces: the direct force exerted by the electric field and the force arising from momentum transfer (the ``electron wind'') from the moving electrons (see Fig.~\ref{Fig:Electrom_fig}a).

Over time, the electron wind can induce metal atoms to migrate in the direction of electron flow, from the cathode to the anode. This migration results in metal atom depletion at the cathode, leading to the formation of voids (Fig.~\ref{Fig:Electrom_fig}b and c) and ultimately causing open circuits. Conversely, metal atoms accumulate at the anode, promoting hillock formation (Fig.~\ref{Fig:Electrom_fig}b) and potentially leading to short circuits.\cite{Black_1969}

The driving force $F_{EM}$, also known as the electron wind force $F_\mathrm{wind}$, governing the electromigration process\cite{Lodder_1998_electromigration} can be expressed as:

\begin{equation}
F_{EM} = Z^\ast e\rho \times j_e,
\end{equation}

\noindent where $\rho$ represents the metal resistivity, $j_e$ denotes the electron current density, $Z^\ast$ is the effective ion valence, and $e$ is the charge of an electron.

Einstein's equation relates the atomic mass flux $J$ to the electron wind force:

\begin{equation}
J = \frac{DC}{k_BT} \times F_{EM},
\end{equation}

\noindent where $C$ is the atom concentration, $D$ is the diffusion constant given by $D = D_0 \exp\left(-\frac{E_A}{k_B T}\right)$, with $D_0$ representing the effective diffusivity along different paths, $E_A$ the activation energy for the dominant diffusion pathway, $k_B$ the Boltzmann constant, and $T$ the temperature.

In a metal line, atoms can diffuse through various pathways: within the bulk of the line, along grain boundaries, and at the interface between the metal and the dielectric. The predominant diffusion path is material-dependent and is determined by its activation energy $E_A$, which, in turn, is governed by the bonding energy of the crystal metal lattice. In Cu interconnects, voids arising from electromigration typically nucleate at the top interface between Cu and the dielectric barrier (typically SiN or SiCN), and void growth subsequently proceeds through grain boundaries.\cite{Hu_1999_Cu,Hau-Riege_2001_Cu, Ogawa_2002_DD_Cu, Hauschildt_2013_IRPS}

One of the critical challenges of downscaling interconnect dimensions is the rapid increase in the relative volume of atoms diffusing along interfaces and grain boundaries, coupled with a decrease in the overall metal volume. This combination leads to a pronounced degradation in the electromigration lifetime of scaled Cu interconnects.\cite{zahedmanesh_2019_Cu, Choi_2018_BEOL_Cu} As a result, the maximum current densities that can be carried reliably (with a 10-year lifetime at 135\deg{}C) steadily decrease in scaled interconnects, imposing increasingly stringent constraints on circuit design and layout.

The reduction in electromigration lifetimes for scaled interconnects has been mitigated (slowed) by the introduction of liner layers between TaN barriers and Cu conductors, particularly at the top interface of the line. Co is currently the primary liner material in use, with Ru emerging as a promising alternative. However, similar to the TaN barrier, which prevents Cu drift into surrounding dielectrics, the thickness of liner layers poses significant challenges for scaling.\cite{witt_testing_2018, Jourdan_2016_Mn_Ru, Pedreira_2019_TaN} Prefill techniques offer a potential alternative to improve reliability, although they introduce additional process complexities.\cite{Pedreira_2018_Co_vias, Pedreira_2022_relia} Analogous to the TaN diffusion barriers, liners contribute minimally to the overall line conductance while occupying an increasing fraction of the line volume. This reduces the available space for the primary conductor (\textit{i.e.}, Cu), leading to a sharp rise in line resistance as interconnect dimensions are scaled down.

An alternative approach to liner scaling is the use of conductor metals with intrinsically high resistance to electromigration, offering significantly better electromigration performance than Cu. Since the activation energy $E_A$ for electromigration generally scales with the cohesive energy of the metal, refractory metals with high melting points present a promising solution. It is noteworthy that metals with high cohesive energies are also advantageous for barrierless dielectric reliability, suggesting that the material properties influencing the different reliability aspects are related. The critical importance of selecting alternative metals with high potential for barrier- and liner-free, reliable metallization schemes is further explored in Sec.~\ref{Sec:Interconnect_Model}.

\subsection{Self-heating and thermal properties}

As previously discussed, the reliability of metal interconnects is heavily influenced by the absolute temperature at critical points within the interconnect structure, as well as the temperature gradients present throughout the metal stack. The temperature in the interconnect is determined by several factors: the thermal resistance of the interconnect structure, the thermal coupling between the interconnects and the heat-generating transistors, self-heating within the interconnect, and the thermal interactions between adjacent metal lines.

As interconnect dimensions continue to shrink and low-$\kappa$ dielectrics are introduced, thermal management becomes increasingly critical. Due to the typically low thermal conductivity of low-$\kappa$ interlayer dielectrics (as low as 0.3 Wm$^{-1}$K$^{-1}$ for OSG 3.0, an organosilicate glass with a dielectric constant of $\kappa = 3.0$),\cite{guralnik_3_2021,oprins_thermal_2024} heat dissipation within the interconnect stack predominantly occurs through metal lines and vias, and is highly dependent on the thermal conductivity of the metals and their connectivity scheme.\cite{chang_thermal_2022} The scaling of metal line width and thickness leads to a reduction in thermal conductivity, an increase in electrical resistivity (see Sec.~\ref{Sec:Resistance}), aggravating self-heating, and an enhanced contribution of barrier thermal resistance.\cite{chang_thermal_2022} As a result, the interconnect stack exhibits higher thermal resistance, with interconnect thermal resistance becoming the dominant factor in overall thermal resistance for advanced packages, leading to increased self-heating of the metal interconnects.\cite{lofrano_joule_2021}

Given that the thermal behavior of interconnect structures is primarily governed by the metallization, accurately predicting the thermal properties of alternative metals is crucial. Interconnect-level thermal models\cite{chang_calibrated_2023,lofrano_towards_2023} can account for factors such as metal line density, via density, and the connectivity between various metal layers. However, these models must also incorporate the size-dependent behavior of the materials to provide a comprehensive understanding of thermal performance.

To accurately capture ballistic thermal transport effects involving both electrons and phonons, \textit{ab initio} simulations can be employed to predict the thermal conductivity of relevant materials. In bulk metals, the total thermal conductivity $K_0$ is the sum of contributions from all electron ($K_\mathrm{el}$) and phonon ($K_\mathrm{ph}$) modes, given by

\begin{equation}
K_0 = K_\mathrm{el} + K_\mathrm{ph} = \sum_\mathbf{k} C_\mathrm{el}\left(\mathbf{k}\right)v^2_\mathrm{el}\left(\mathbf{k}\right)\tau_\mathrm{el}\left(\mathbf{k}\right) + \sum_\mathbf{q} C_\mathrm{ph}\left(\mathbf{q}\right)v^2_\mathrm{ph}\left(\mathbf{q}\right)\tau_\mathrm{ph}\left(\mathbf{q}\right)
\end{equation}

\noindent Here, $C$ represents the heat capacity, $v$ is the group velocity, and $\tau$ denotes the relaxation time. The subscripts `el' and `ph' refer to the contributions of electrons and phonons, respectively, while $\mathbf{k}$ and $\mathbf{q}$ represent the wavevectors of electrons and phonons, respectively. 

In semiconductors, heat is transported primarily by phonons, whereas in metals, electrons are the dominant heat carriers. Consequently, in metals, the thermal conductivity $K_0$ and the electrical resistivity $\rho_0$ are related through the Wiedemann--Franz law

\begin{equation}
K_0 = \frac {LT}{\rho_0 },
\end{equation}

\noindent where $L$ denotes the Lorenz number and $T$ the temperature. For free electrons, $L = 2.4\times 10^{-8}$ W$\Omega$K$^{-2}$. Many bulk metals exhibit Lorenz numbers close to this free-electron value (\textit{e.g.}, Cu\cite{hust_lorenz_1973}), although certain metals, such as those in the Pt group\cite{powell_thermal_1962} or W,\cite{hust_lorenz_1973} show Lorenz numbers approximately 10 to 15\%{} higher. Such deviations from free electron behavior typically stem from additional contributions of phonon transport to the thermal conductivity.

\begin{figure}[tb]
\includegraphics[width=16cm]{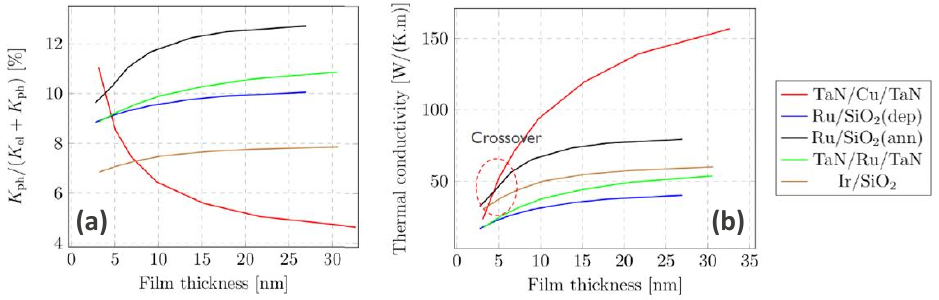} 
\caption{\label{Fig:Therm_cond} Thickness-dependent thermal conductivity: (a) Phonon-dominated regime, (b) Total thermal conductivity including contributions of both phonons and electrons.}
\end{figure}

To account for the influence of reduced metal dimensions on thermal conductivity, a thin-film model based on Mayadas--Shatzkes framework\cite{mayadas_electrical-resistivity_1970} (see Sec.~\ref{Sec:Semiclassical}) has been employed to estimate the size-dependent thermal conductivity for various metals (or metal stacks). Experimentally calibrated reflection coefficients and surface specularity parameters (see Sec.~\ref{Sec:Downselection}) were used. This model allows for the separation of phonon and electron contributions to total thermal conductivity.

As shown in Fig.~\ref{Fig:Therm_cond}a, phonon contributions remain below 15\%{} for relevant interconnect dimensions, indicating electron-dominated thermal transport and (approximate) applicability of the Wiedemann-Franz law. Figure~\ref{Fig:Therm_cond}b illustrates the thickness dependence of the thermal conductivity for several metals. Size-dependent thermal conductivity behavior is influenced by electron mean free path, grain boundary scattering, and surface scattering characteristics.\cite{cahill_nanoscale_2002,cahill_nanoscale_2014} The extent to which the Lorenz number is modified in nanoscale metal films and wires is an ongoing area of research. While theory and experiments suggest a possible reduction from bulk values, conclusive evidence is still lacking.\cite{li_thermal_2003,stojanovic_thermal_2010,wang_breakdown_2013,huang_thermal_2014,sawtelle_temperature-dependent_2019}

\section{Alternative metal screening and downselection\label{Sec:Downselection}}

Due to the complexities and high costs associated with fabricating nanoscale interconnect lines at target dimensions, direct experimental identification of of the most auspicious metals among numerous candidates is impractical. Therefore, a simplified procedure needs to be devised to identify and select the most promising candidates based on easily measurable parameters for a wide range of materials. While the specific set of criteria may vary, recent approaches have focused on both nanoscale metal resistivity and reliability.\cite{adelmann_alternative_2014,sankaran_exploring_2014-1,sankaran_exploring_2014} Further insights into candidate metal properties can be gained through thin-film experiments, which can serve as initial approximations of expected line performance, following the workflow in Fig.~\ref{fig:workflow}. The following sections discuss the current state of understanding for elemental, binary, and ternary metals based on this workflow.

\subsection{\label{Sec:Elements_Proxies}Elemental metals}

Considering the discussions in Secs.~\ref{Sec:Resistance} and \ref{Sec:Reliability}, the following three material properties have been identified as representative indicators of the overall performance of metals in scaled interconnects.

\begin{enumerate}[(i)]
\item the bulk resistivity, $\rho_0$; 
\item the mean free path of the charge carriers $\lambda$ or, alternatively, the product of the bulk resistivity and the mean free path of the charge carriers, $\rho_0\times\lambda$; 
\item the cohesive energy or, alternatively, the melting temperature.
\end{enumerate}

\noindent The first two indicators represent the potential for low resistivity at small dimensions, as discussed in Sec.~\ref{Sec:Resistance}. The third parameter can be considered an indicator for electromigration resistance and barrierless reliability, as more refractory metals generally exhibit better performance (see Sec.~\ref{Sec:Reliability}).

\begin{figure}[tb]
\includegraphics[width=16cm]{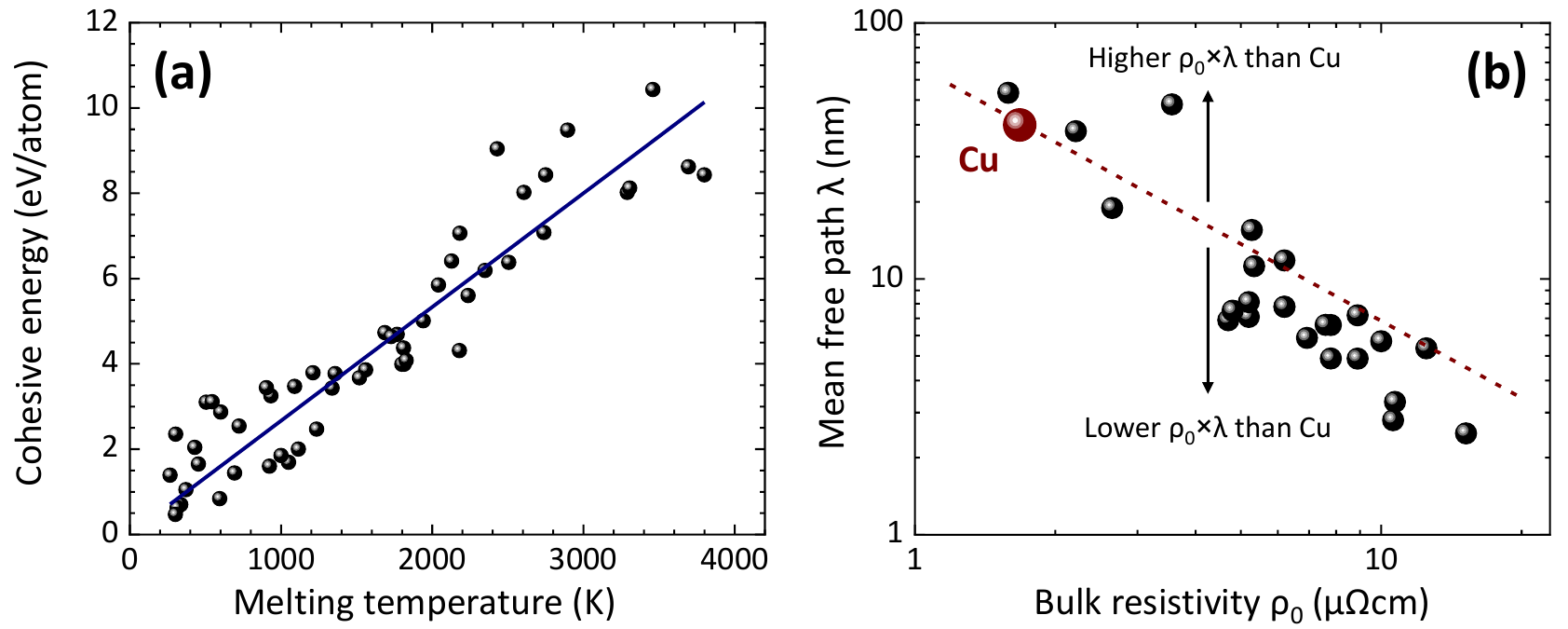} 
\caption{\label{Fig:EcohMFP} (a) Relationship between calculated cohesive energy and melting temperature for various elements. (b) Comparison of calculated mean free path $\lambda$ from the $\rho_0\times\lambda$ product and experimental resistivities $\rho_0$ for elemental metals with resistivities below 20 $\mu\Omega$cm. Cu is highlighted for reference, with the dotted line representing its expected trend for the Cu $\rho_0\times\lambda = 6.8 \times 10^{-16}$ $\Omega$m).}
\end{figure}

Values for the first indicator (i) can be found in the literature. While \textit{ab initio} calculations of $\rho_0$ are feasible, they are computationally intensive and unsuitable for screening a large number of metals. Similarly, calculating the mean free path $\lambda$ is resource-intensive. However, the product $\rho_0\times\lambda$ (or the related tensor components) can be obtained relatively easily from \textit{ab initio} calculations, as explained in Sec.~\ref{Subsec:Rholambda}. The third proxy can be either obtained from the literature (melting point) or calculated (cohesive energy). A strong correlation between melting points and cohesive energies has been observed (Fig.~\ref{Fig:EcohMFP}a), justifying their interchangeable use.\cite{sankaran_exploring_2014-1,sankaran_exploring_2014}

In practice, the product of bulk resistivity and mean free path, $\rho_0\times\lambda$ has been preferred as a figure of merit for metals. However, it's essential to consider parameter (i), the bulk resistivity $\rho_0$, in conjunction with $\rho_0\times\lambda$. While this is straightforward for elemental metals, it can be challenging for binary and ternary metals due to limited knowledge of their fundamental properties.

\begin{figure}[tb]
\includegraphics[width=10cm]{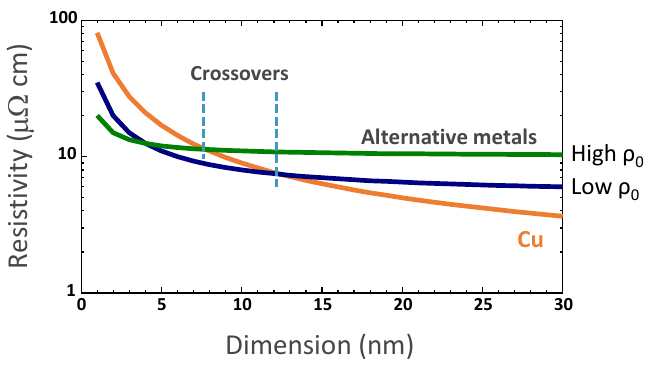} 
\caption{\label{Fig:Crossover} Resistivity scaling trends: metals with small mean free paths $\lambda$ may have comparable or lower resistivity than Cu at reduced dimensions. The crossover dimension depends however on the bulk resistivity $\rho_0$ for a constant $\rho_0\times\lambda$ product.}
\end{figure}

The primary limitation of using the $\rho_0\times\lambda$ figure of merit alone is the inherent correlation between $\rho_0$ and $\lambda$: short mean free paths (short relaxation times) lead to high resistivities, as evident from Eq.~\eqref{eq:sigma}. Figure~\ref{Fig:EcohMFP}b demonstrates this relationship by plotting the mean free path deduced from the $\rho_0\times\lambda$ product against bulk resistivity $\rho_0$ for various elemental metals. As shown in Fig.~\ref{Fig:Crossover}, a short $\lambda$ can result in a less pronounced thickness-dependent resistivity compared to Cu, leading to a resistivity crossover at a finite dimension. However, even with a constant $\rho_0\times\lambda$, a lower bulk resistivity $\rho_0$ in an alternative metal is still advantageous, as it typically results in a crossover with Cu resistivity at larger dimensions compared to metals with higher $\rho_0$ (see Fig.~\ref{Fig:Crossover}). While metals with the smallest $\lambda$ may eventually exhibit the lowest resistivity, the crossover with Cu resistivity might occur at dimensions that are not practical for interconnect applications.

The proposed indicators can be applied to all metallic elements in the periodic table. Using the $\rho_0\times\lambda$ product of Cu ($\rho_0\times\lambda = 6.8 \times 10^{-16}$ $\Omega$m), a bulk resistivity of 10 $\mu\Omega$cm, and the melting point of Cu (1358 K) as cutoff values, the most promising elemental metals can be identified, as shown in Tab.~\ref{Tab:Elements}. For reference, the properties of Cu are also included. The list includes several transition metals, among them Pt-group metals like Ru, Rh, and Ir.

\begin{table}
\caption{\label{Tab:Elements}Properties of prospective alternative metals (and Cu for reference): crystal structure, bulk resistivity $\rho_0$, calculated $\rho_0\times\lambda$ figure of merit (see Sec.~\ref{Subsec:Rholambda}), mean free path $\lambda$, melting temperature, and cohesive energy.\cite{gall_electron_2016, dutta_thickness_2017, bass_1.2._1983, haynes_crc_2014}}
    \centering
    \begin{tabular}{m{1.3 cm}>{\centering}m{1.8 cm}>{\centering}m{2.7 cm}>{\centering}m{2.1 cm}>{\centering}m{2.2 cm}>{\centering}m{2.3 cm}>{\centering\arraybackslash}m{2.3 cm}}
\toprule
& Crystal & Bulk resistivity & $\rho_0\times\lambda$ & Mean free & Melting  & Cohesive \\
& structure & $\rho_0$ ($\mu\Omega$cm) & 10$^{-16}$ $\Omega$m & path $\lambda$ (nm) & temp. (K) & energy (eV) \\
\hline 
Cu &  fcc & 1.68 & 6.8 & 40.7 & 1358 & 3.8 \\
\multirow{2}{*}{Co} &  \multirow{2}{*}{hcp} & $zz,\, \parallel$ 5.1 & 4.8 & 9.4 & \multirow{2}{*}{1768} & \multirow{2}{*}{4.7} \\
& & $(xx = yy,\,\perp)$ 9.1 & 7.5 & 8.2 & & \\
Ni &  fcc & 6.93 & 4.1 & 5.9 & 1728 & 4.6 \\
Mo &  bcc & 5.3 & 5.8 & 10.9 & 2895 & 9.5 \\
\multirow{2}{*}{Ru} &  \multirow{2}{*}{hcp} & $zz,\, \parallel$ 5.7 & 3.8 & 6.7 & \multirow{2}{*}{2606} & \multirow{2}{*}{8.0} \\
& & $(xx = yy,\, \perp)$ 7.4 & 5.1 & 6.9 & & \\
Rh &  fcc & 4.7 & 3.2 & 6.8 & 2236 & 5.6 \\
Ir &  fcc & 5.0 & 3.8 & 7.5 & 2719 & 7.1 \\
\toprule
\end{tabular}
\end{table}

\begin{figure}
\includegraphics[width=10cm]{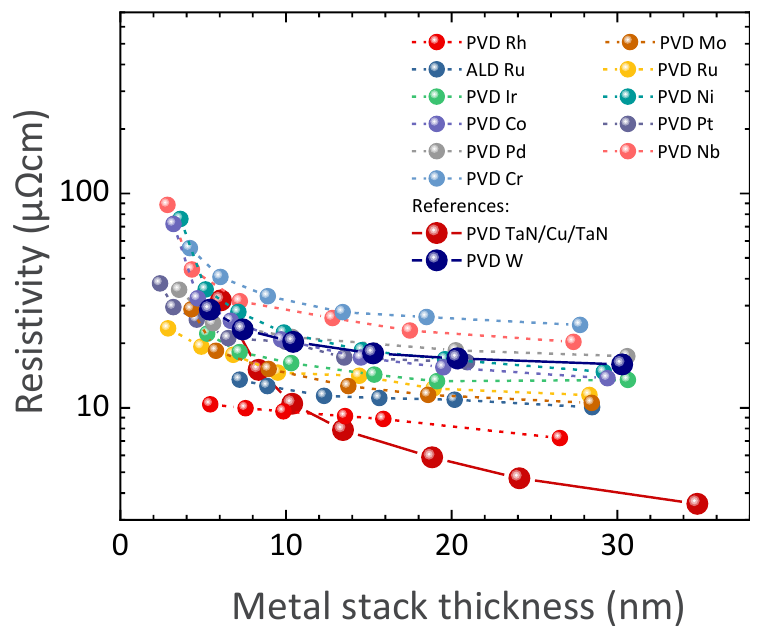} 
\caption{\label{fig:el_res} Experimental resistivity scaling trends for thin film metals: thin film resistivities \textit{vs.}\ metal film (stack) thickness for various potential alternative metals. Pt-group data from  Refs.~\onlinecite{dutta_thickness_2017,popovici_atomic_2017}.}
\end{figure}

The limited number of promising elements allows for experimental thin-film studies on all candidates, following the workflow outlined in Fig.~\ref{fig:workflow}. Figure~\ref{fig:el_res} illustrates the thickness-dependent resistivity of various elemental metal thin films. As expected due to its long mean free path, Cu exhibits a rapid increase in resistivity with decreasing thickness, particularly below 10 nm. In contrast, many alternative elemental metals demonstrate a much less pronounced increase, resulting in comparable or even lower thin-film resistivities at thicknesses below 5 to 10 nm, depending on the metal. Notably, several Pt-group metals, including Rh, Ir, and Ru, exhibit low resistivities, making them promising candidates for future interconnect applications. Resistivity modeling\cite{dutta_thickness_2017} has confirmed that the weaker film thickness dependence and ultimately lower thin-film resistance compared to Cu can be attributed to shorter mean free paths.

Another promising metal candidate is Mo,\cite{adelmann_alternative_2014,gall_electron_2016, gall_metals_2018,gall_search_2020,founta_properties_2022} which, despite exhibiting slightly higher thin-film resistivity, offers several advantages for interconnect integration and is significantly more cost-effective than the expensive Pt-group metals. Consequently,  Mo has emerged as a potential candidate for both logic and memory interconnects.\cite{founta_properties_2022,gupta_barrierless_2022,hosseini_ald_2022,gupta_buried_2021}

\begin{figure}
\includegraphics[width=8.5cm]{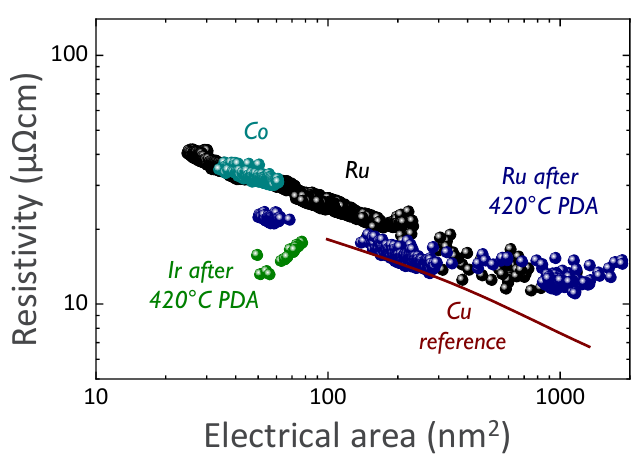} 
\caption{\label{fig:ms5} Resistivity scaling of narrow interconnect lines: comparison of Co, Ru, and Ir data obtained from metallic spacer etched test structures.\cite{dutta_highly_2017,dutta_ruthenium_2017,dutta_finite_2018,dutta_sub-100_2018,adelmann_alternative_2018} Co and Ru shown data are for as-deposited metals, while Ir and additional Ru data were obtained after are post-deposition annealing at 420\deg{}C. A reference line is shown for Cu dual-damascene metallization.\cite{ciofi_impact_2016}}
\end{figure}

A metal-spacer-defined metal-etch short loop vehicle, developed at imec, has provided additional insights into the resistivity of ultrasmall nanowires and expected line resistance scaling. This vehicle enables the fabrication of nanowires with cross-sectional areas below 100 nm$^2$.\cite{dutta_highly_2017,dutta_ruthenium_2017,dutta_finite_2018,dutta_sub-100_2018} The results in Fig.~\ref{fig:ms5} demonstrate that Ir interconnects exhibit superior resistivity scaling compared to state-of-the-art Cu damascene interconnects, while Ru and Co (to a lesser extent) show similar resistivities for sufficiently narrow lines. These results confirm the potential of these materials as alternatives to Cu for interconnect metallization. It is noteworthy that, despite comparable resistivities, Ru can offer significantly better line resistance than Cu when integrated without barriers and liners due to its larger conducting volume. This aspect will be discussed further in Sec.~\ref{Sec:Interconnect_Model}.

Beyond thin film and short-loop nanowire resistivity evaluations, selected alternative metals have been integrated into scaled interconnect lines to assess their line resistance scalability as well as their reliability, particularly with regard to the need for barrier layers. Specific results are now available for Co, Ru, and Mo.

Co metallization has been the subject of extensive recent research and has been integrated into commercial high-volume circuit manufacturing.\cite{auth_10nm_2017,yeoh_interconnect_2018,griggio_reliability_2018} However, Co migrates into surrounding low-$\kappa$ dielectrics, requiring a barrier layer (typically TiN) analogous to Cu.\cite{griggio_reliability_2018} In advanced technology nodes with ultrathin low-$\kappa$ films, maintaining hydrophobic interfaces and a continuous thick barriers are however crucial to prevent metal drift, posing significant challenges to the integration and scalability of Co metallization in advanced interconnects.\cite{tierno_cobalt_2019, tierno_impact_2020}

In contrast to Co, Mo has emerged as a promising candidate for barrierless integration, eliminating the need for an adhesion layer or diffusion barrier on SiO$_2$, low-$\kappa$ organosilicate glasses, or SiCO dielectric films. This finding has been corroborated by multiple studies.\cite{lesniewska_dielectric_2020, lesniewska_reliability_2021}  

Ru is another potential barrierless alternative to Cu. Its high cohesive energy and excellent oxidation resistance allow for barrierless integration, although a thin adhesion-promoting layer might be necessary. However, this adhesion layer (typically TiN) can be scaled down to a few \AA{}, well below the minimum thickness required for functional diffusion barriers.\cite{wen_atomic_2016} Further research has confirmed that the reliability of Ru is not compromised when integrated with a 0.3 nm TiN adhesion layer, with no evidence of metal drift. Additionally, Ru combined with dense low-$\kappa$ dielectrics ($\kappa = 3.0$) has demonstrated a 10-year lifetime based on damascene time-dependent dielectric breakdown  results.\cite{hosseini_ald_2022, tierno_reliability_2021} 

\begin{figure}[tb]
\includegraphics[width=14cm]{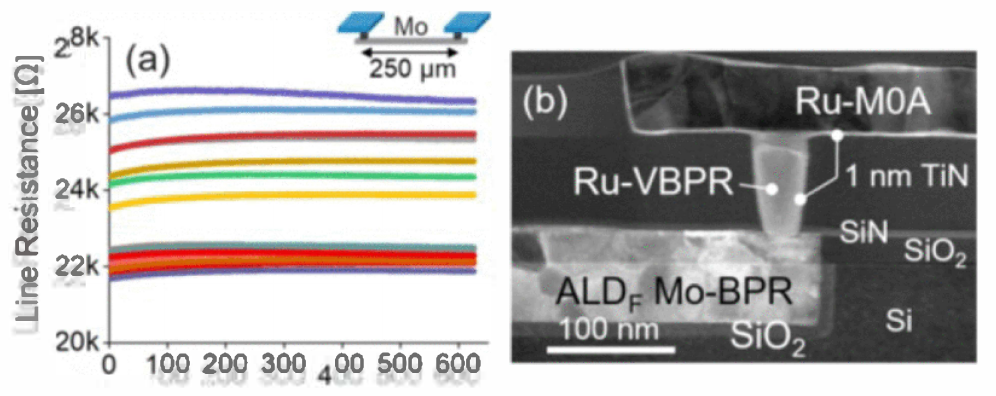} 
\caption{\label{Fig:Electrom_fig4} Electromigration stability of Mo-Ru hybrid interconnects: (a) No failures were observed after more than 600 hours testing at 330\deg C and 5 MA/cm$^2$. (b) Cross-sectional transmission-electron micrograph of the tested structure. Reprinted with permission from Ref.~\onlinecite{gupta_barrierless_2022}.}
\end{figure}

Both Mo and Ru interconnects exhibit exceptional electromigration performance. As a matter of fact, observing electromigration failures in Mo and Ru interconnects has been challenging due to the extreme temperatures and current densities required.\cite{gupta_barrierless_2022, Pedreira_2017_Co_Ru} Figure~\ref{Fig:Electrom_fig4} illustrates an example of electromigration testing for Mo interconnects, where no failures were observed even after more than 600 hours of stressing at 330\deg C.\cite{gupta_barrierless_2022}

\begin{figure}[tb]
\includegraphics[width=14cm]{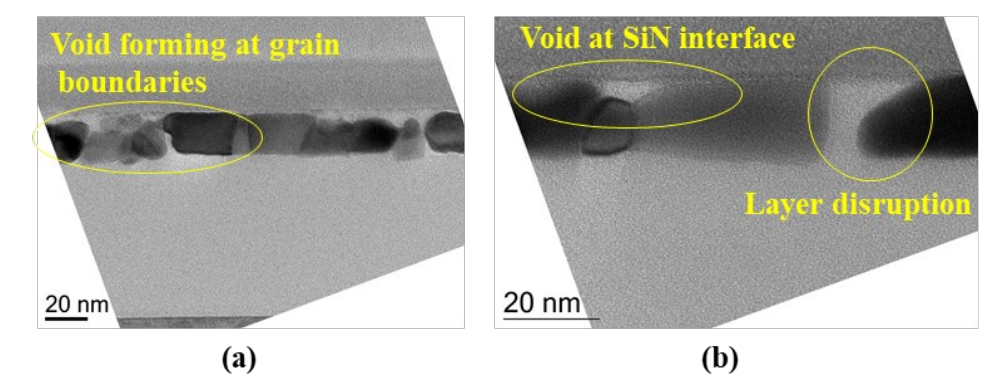} 
\caption{\label{Fig:Electrom_fig5} Transmission electron micrographs illustrating electromigration-induced void formation in Ru interconnects: voids located (a) at grain boundaries and (b) at the dielectric interface. Reprinted with permission from Ref. \onlinecite{Pedreira_2020_relia}.}
\end{figure}

Furthermore, Hu \textit{et al.}\cite{hu_future_2018} demonstrated excellent reliability in 36 nm wide Ru interconnect lines. Additionally, Varela Pedreira \textit{et al.}\cite{Pedreira_2017_Co_Ru} reported unsuccessful attempts to induce electromigration failures in barrierless 21 nm metal pitch Ru interconnects, even at high current densities exceeding 30 MA/cm$^2$. In the same study, the authors subjected barrierless Ru lines to even more extreme conditions, driving current densities of 150 to 200 MA/cm$^{2}$, which resulted in significant self-heating ($\sim 260$\deg C). Only under these extreme conditions, void formation was observed at the grain boundaries and the dielectric interface (Fig.~\ref{Fig:Electrom_fig5}).\cite{Pedreira_2020_relia} Moreover, Beyne \textit{et al.}\cite{Beyne_2019_ActivationE} investigated scaled Ru wires with rough surfaces and no barrier using low-frequency noise measurements. Their findings suggest that the metal/dielectric interface serves as the primary diffusion path.

The results of the combined modeling and experimental screening process for elemental metals, following the workflow outlined in Fig.~\ref{fig:workflow}, are summarized in Tab.~\ref{Tab:Elements}. Six elemental metals have emerged as promising candidates for advanced interconnect metallization: the platinum-group metals Rh, Ru, and Ir, along with the transition metals Ni, Co, and Mo. Among these, Co has already been integrated into local interconnects of commercial CMOS technologies.\cite{auth_10nm_2017,yeoh_interconnect_2018,griggio_reliability_2018} Mo and Ru have garnered the most attention for use in highly scaled lines and vias, due to their demonstrated reliability, even in barrierless configurations. Currently, barrierless Ru metallization combined with airgap structures is considered the leading candidate for logic interconnects with sub-20 nm pitches,\cite{zhang_ruthenium_2016,tokei_on_chip_2016,wen_ruthenium_2016,wan_subtractive_2018,tokei_inflection_2020,nogami_interconnect_2023} while Mo is also being actively investigated as a viable alternative, particularly for memory applications.\cite{tokei_inflection_2020,gupta_buried_2021,hosseini_ald_2022,gupta_barrierless_2022,nogami_interconnect_2023}

\subsection{Graphene and hybrid graphene-metal metallization \label{Sec:Hybrid}}

In addition to conventional metals, graphene has also been proposed as a potential conductor for advanced interconnects.\cite{murali_resistivity_2009,rakheja_evaluation_2013,jiang_cmos-compatible_2018,jiang_ultimate_2019} Graphene exhibits high carrier mobility and can support large current densities while demonstrating excellent resistance to electromigration.\cite{jiang_characterization_2017,agashiwala_reliability_2020} However, pristine graphene is a semimetal with a low charge carrier density, resulting in high sheet resistance, which limits its direct applicability in interconnects.\cite{xu_graphene_2008} To overcome this, multilayer graphene structures and doping are required to reduce resistivity to practical levels. 

\begin{figure}[tb]
\includegraphics[width=8cm]{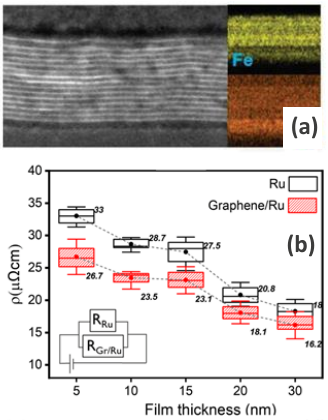} 
\caption{\label{Fig:Graphene} (a) Cross-sectional transmission electron micrograph and energy-dispersive x-ray spectroscopy chemical mapping of FeCl$_3$-intercalated graphene. Reprinted with permission from Ref.~\onlinecite{li_intercalated_2023}. (b) Resistivity measurements of bare and graphene-capped Ru interconnects for varying Ru thicknesses. Reprinted with permission from Ref.~\onlinecite{achra_metal_2021}.}
\end{figure}

One effective doping method is intercalation (see Fig.~\ref{Fig:Graphene}a), which has demonstrated considerable lowering of thin film resistivities.\cite{xu_modeling_2009,agashiwala_demonstration_2021} Notably, FeCl$_3$-intercalated graphene has recently achieved resistivity values comparable to, or even lower than, those of Cu.\cite{khrapach_novel_2012,jiang_intercalation_2017,li_intercalated_2023,huang_intercalated_2023} The intercalation process preserves the integrity of the graphene Dirac cone while modulating the Fermi level, thereby increasing the charge carrier concentration. Nevertheless, despite the much improved resistivity in doped intercalated graphene, the high contact resistance when co-integrated with metallic vias remains a significant challenge. Future advancements may arise from exploring alternative $n$-type intercalation species,\cite{li_intercalated_2023} but further research is necessary to demonstrate the successful integration of intercalated multilayer graphene into scaled interconnect architectures.

In addition to intercalated graphene, graphene--metal hybrid composite metallization is another promising approach for interconnects. For instance, studies of Ru thin films capped with multilayer graphene have demonstrated a reduction in both sheet resistance and effective resistivity by approximately 10 to 20\%{} compared to uncapped Ru films (see Fig.~\ref{Fig:Graphene}b).\cite{achra_metal_2021} This improvement has been attributed to a 0.5 eV reduction in the Fermi level, indicative of $p$-type doping in the graphene layer, likely driven by charge transfer from Ru.\cite{achra_characterization_2020} Similar reductions in sheet and line resistance have been observed in Cu--graphene hybrid metallization schemes as well.\cite{mehta_enhanced_2015,son_copper-graphene_2021} While many studies have focused on hybrids incorporating single-layer graphene, the use of bilayer or multilayer graphene may yield even greater resistance reductions, although the potential benefits could be limited by charge screening effects and interlayer resistance.

Beyond thin film studies, the integration of graphene--metal hybrid composite materials presents several challenges. The deposition temperatures required to produce high-quality graphene typically exceed the thermal budget for interconnect processing, which is constrained to around 400\deg{}C. Therefore, the development of low-temperature processes for defect-free graphene deposition is essential.\cite{li_low-temperature_2011} Moreover, when integrated, \textit{e.g.}, in a metal patterning scheme, graphene should ideally be deposited selectively on the sidewalls of patterned interconnect lines to avoid shorts between adjacent lines. Alternatively, graphene integration within damascene interconnect architectures has also been explored,\cite{kang_evaluation_2020} resulting in substantial improvements in electromigration resistance.\cite{nogami_electromigration_2021} However, further research is needed to validate the performance benefits of such hybrid systems in realistic scaled interconnect structures.

\subsection{Binary intermetallics\label{Sec:binary}}

The preceding analysis of elemental metals can be considered exhaustive, encompassing all metals of potential interest. To expand the range of materials of potential interest, recent research has focused increasingly on compound metals. Among binary or ternary metal systems, disordered compounds (alloys) typically exhibit high resistivities due to strong alloy scattering (see Sec.~\ref{Sec:Resistivity_Intermet}). In contrast, numerous ordered intermetallic systems have demonstrated experimentally low resistivities, typically within a specific narrow composition range.\cite{terada_thermal_2002} 

Figure~\ref{fig:alloy_intermetallic} illustrates a well-known example of the Au--Cu system,\cite{Johansson_CuAu_1936} which includes the intermetallics AuCu$_3$ and AuCu. The figure shows that the resistivities of these intermetallics can be significantly lower than those of random alloys within the same material system and may even approach the resistivities of the constituent elemental metals.

To benchmark and downselect binary intermetallics, the criteria discussed for elemental metals in Sec.~\ref{Sec:Elements_Proxies} can be applied equally. However, a comprehensive benchmarking and downselection process, similar to that conducted for elemental metals, remains beyond reach due to the vast number of potential intermetallic compounds and the limited available data for many of them. Nevertheless, the $\rho_0\times\lambda$ figure of merit can be calculated for selected intermetallics using \textit{ab initio} methods, as described in Sec.~\ref{Subsec:Rholambda}. It is important to note that many intermetallics possess non-cubic crystal structures, often requiring tensor formulations. Melting points can generally be found in the literature, such as, \textit{e.g.}, from binary phase diagrams, or replaced by calculated cohesive energies when not readily available. 

\begin{figure}[tb]
\includegraphics[width=10cm]{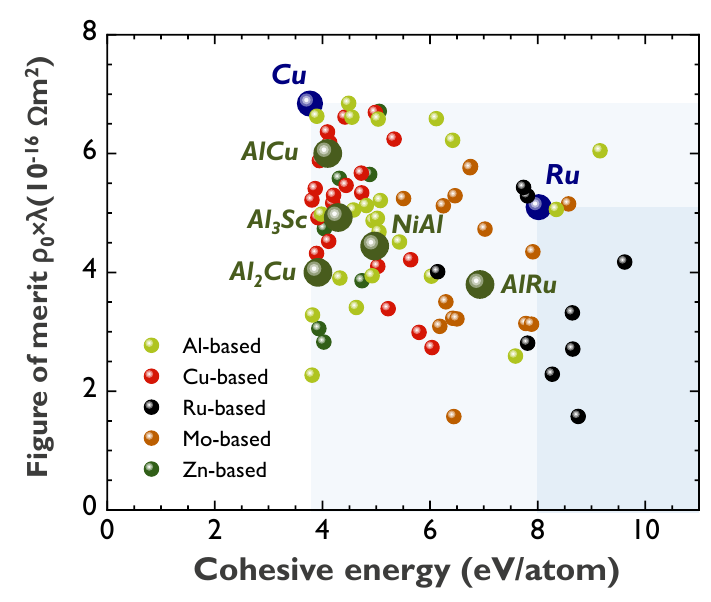} 
\caption{\label{Fig:Intermetallic_screening} \textit{Ab initio} screening of binary intermetallics: Calculated $\rho_0\times\lambda$ product \textit{vs.}\ cohesive energy. Several Al-based intermetallics, highlighted by larger dots, show promising properties based on the screening criteria described in the text. Cu and Ru are added as reference.}
\end{figure}

Thus far, \textit{ab initio} screening studies have identified numerous binary intermetallics with low $\rho_0\times\lambda$ values and cohesive energies exceeding that of Cu (Fig.~\ref{Fig:Intermetallic_screening}); however, only few surpass Ru. A significant obstacle in evaluating intermetallics is the frequent absence of accurate reported bulk resistivities $\rho_0$. As a result, a downselection process as comprehensive as that for elemental metals is not readily achievable.

Experimental investigations have primarily focused on several aluminide intermetallics due to their low bulk resistivities (below 10 $\mu\Omega$cm), higher melting points compared to Cu, and notable oxidation resistance.\cite{howell_developing_2011} Among the most promising candidates are NiAl, AlCu, Al$_2$Cu, AlRu, and Al$_3$Sc. The properties of these aluminide intermetallics are summarized in Tab.~\ref{Tab:Aluminides}, and can be directly compared to those of the promising elemental metals listed in Tab.~\ref{Tab:Elements}. For these aluminides, experimental studies have demonstrated low resistivities,\cite{soulie_thickness_2020, chen_cual2_2019} although this has predominantly been observed for relatively thick films with thicknesses $\gg 10$ nm.

Among the most promising aluminide intermetallics, NiAl has emerged as the most extensively studied compound.\cite{chen_nial_2018,soulie_thickness_2020,soulie_aluminide_2021,soulie_improved_2022, chen_liner_2019} In physical vapor-deposited films on 300 mm Si substrates, a resistivity of 13.9 $\mu\Omega$cm was achieved at a film thickness of 56 nm after post-deposition annealing at 600\deg C.\cite{soulie_thickness_2020} This resistivity can be further reduced by depositing NiAl at an elevated temperature of 420\deg C, followed by an \textit{in situ} Si capping layer to prevent surface oxidation. Under these conditions, a resistivity of 18 $\mu\Omega$cm was obtained for a 22 nm thick film.\cite{soulie_improved_2022} To achieve even lower resistivities at reduced thicknesses, epitaxial NiAl films on Ge (100) have been explored, resulting in a resistivity as low as 11.5 $\mu\Omega$cm for a 7.7 nm thick film.\cite{soulie_iitc_2023} However, integrating such epitaxial layers into scalable and manufacturable interconnects remains a substantial challenge.

AlCu and Al$_2$Cu films\cite{chen_cual2_2019,soulie_aluminide_2021, Kuge2022, Kuge2023,soulie_AlCu_2024} with thicknesses around 10 nm have demonstrated resistivities below 20 $\mu\Omega$cm, and below 10 $\mu\Omega$cm for films around 30 nm in thickness, after post-deposition annealing at 500\deg C. The resistivity of Al$_2$Cu is lower than that of Ru for film thicknesses of 10 nm and above, while both AlCu and Al$_2$Cu outperform Mo across the entire studied thickness range from 5 to 30 nm. Additionally, these compounds exhibit resistivities comparable to TaN/Cu/TaN for thicknesses below 8 nm. Al$_2$Cu also displays excellent gap-filling capabilities and promising reliability metrics in time-dependent dielectric breakdown, electromigration, and bias temperature stress tests.\cite{koike_intermetallic_2021} However, further investigation is required to fully validate the potential of AlCu and Al$_2$Cu for advanced interconnects with high reliability.

A resistivity of 12.6 $\mu\Omega$cm has been reported for a 24 nm Al$_3$Sc thin film following post-deposition annealing at 500\deg C.\cite{soulie_al3sc_2024} The resistivity was limited by a combination of grain boundary scattering and point defect (disorder) scattering, which presents significant challenges for further improvements. AlRu has also been identified as a potential candidate to replace Cu; however, experimental resistivities for AlRu have remained comparatively high thus far compared to those of the other aluminides discussed here, primarily due to challenges in producing highly ordered films with large grains.\cite{fang_AlRu_2024}

For Cu$_2$Mg, a resistivity of 25.5 $\mu\Omega$cm has been reported for a 5 nm thick film, along with excellent gap-filling performance achieved via sputtering reflow. However, a thick MgO layer formed within the underlying SiO$_2$ due to interfacial reactions between Cu$_2$Mg and SiO$_2$. This interfacial reaction raises significant concerns regarding the feasibility of integrating Cu$_2$Mg into scaled interconnects, where (near-)zero interface formation is essential to achieve low-resistance lines, casting doubt on its suitability for such applications.\cite{Chen_iitc_2020}

\begin{table}
\caption{\label{Tab:Aluminides} Properties of prospective intermetallics as alternative metals, including Cu as a reference: crystal structure, bulk resistivity, calculated $\rho_0 \times \lambda$ figure of merit (see Sec.~\ref{Subsec:Rholambda}), deduced mean free path $\lambda$, melting temperature, and calculated cohesive energy of various intermetallics. Note that AlCu and Al$_2$Cu exhibit transitions to different high-temperature phases between 850 K and 900 K, rather than to the liquidus.\cite{zobac_experimental_2019}}
\centering
\begin{tabular}{m{1.3 cm}>{\centering}m{2.0 cm}>{\centering}m{2.7 cm}>{\centering}m{2.1 cm}>{\centering}m{2.2 cm}>{\centering}m{2.3 cm}>{\centering\arraybackslash}m{2.3 cm}}
\toprule
& Crystal & Bulk resistivity & $\rho_0\times\lambda$ & Mean free & Melting  & Cohesive \\
& structure & $\rho_0$ ($\mu\Omega$cm) & 10$^{-16}$ $\Omega$m & path $\lambda$ (nm) & temp. (K) & energy (eV) \\
\hline 
Cu &  fcc & 1.68 & 6.8 & 40.7 & 1358 & 3.8 \\
AlNi \cite{chen_nial_2018} &  Pm$\overline{3}$m (B2) & 5.5 & 4.4  & 8  & 1910 &  5.0 \\
AlRu \cite{fang_AlRu_2024} &  Pm$\overline{3}$m (B2) & $\sim 10$ &  3.8 & $\sim$4  & 2250 & 6.9  \\
AlCu \cite{Kuge2023} &  C2/m & 8 & 6.0  & 7.5 & N/A & 4.1  \\
Al$_2$Cu \cite{chen_cual2_2019} &  Fm$\overline{3}$m & 6.5 &  4.0 & 5.5 & N/A &  3.9 \\
Al$_3$Sc \cite{soulie_al3sc_2024} &  Pm$\overline{3}$m (L1$_2$) & 7 & 4.9 & 7 & 1280 & 4.3 \\
Cu$_2$Mg \cite{chen_interdiffusion_2021} &  Fd$\overline{3}$m & 5.7  &  9.6 &  22  & 1073 &  2.9 \\
CuTi \cite{CuTi_2023} &  P4/nmm & 19.5  & 3.4 & 12.5 & 1260 & 4.3 \\
\toprule
\end{tabular}
\end{table}

\begin{figure}[tb]
\includegraphics[width=14cm]{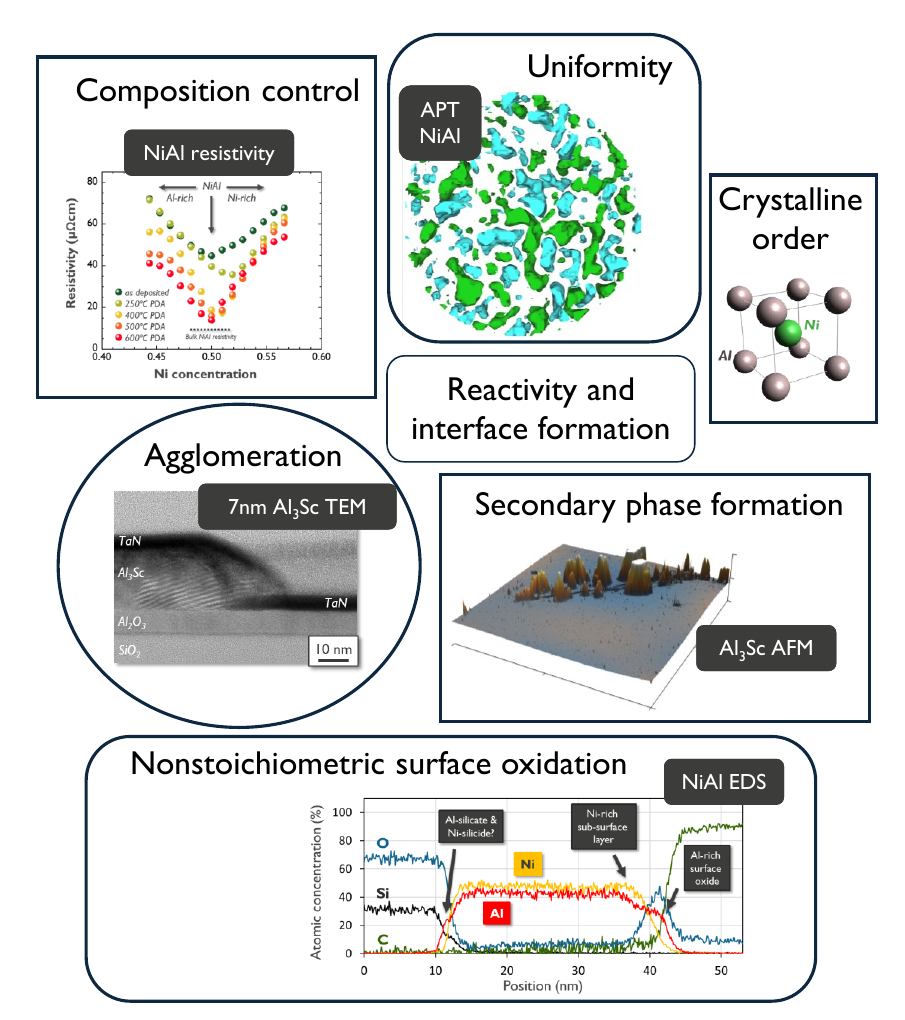} 
\caption{\label{Fig:ChallengeAl} Common challenges for obtaining low resistivties in intermetallic compounds: composition dependence of resistivity, nanoscale composition uniformity, secondary phase formation, crystalline order, reactivity and interface formation, agglomeration, as well as nonstoichiometric surface oxidation.}
\end{figure}

\begin{figure}[tb]
\includegraphics[width=8.2cm]{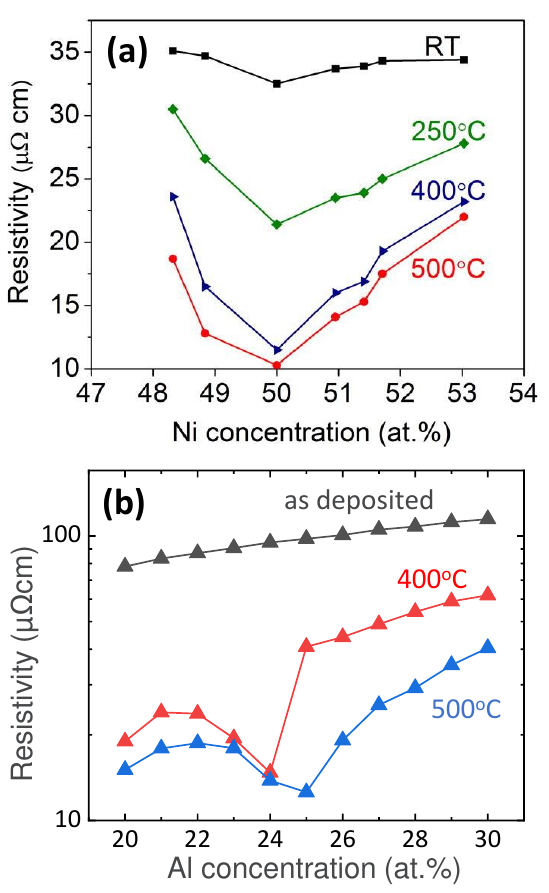} 
\caption{\label{Fig:Resistivity} (a) Resistivity of Ni$_{x}$Al$_{1-x}$ \textit{vs.}\ Ni concentration around stoichiometric NiAl. Reprinted with permission from Ref.~\onlinecite{chen_nial_2018}. (b) Resistivity of Al$_x$Sc$_{1-x}$ \textit{vs.}\ Al concentration around stoichiometric Al$_3$Sc.\cite{soulie_aluminide_2021,soulie_al3sc_2024} In both cases, a pronounced resistivity minimum is observed at the stoichiometric composition.}
\end{figure}

In contrast to elemental metals, binary intermetallics present several additional challenges, including crystalline order and the minimization of point defect densities, precise control of composition and its uniformity, the formation of secondary phases, agglomeration, (interface) reactivity, and nonstoichiometric surface oxidation, as exemplified in Fig.~\ref{Fig:ChallengeAl}. A primary challenge lies in controlling the composition of binary intermetallics, as reported for the Al$_{1-x}$Ni$_{x}$ and Al$_x$Sc$_{1-x}$ systems.\cite{chen_nial_2018, soulie_al3sc_2024} As shown in Fig.~\ref{Fig:Resistivity}a, the resistivity of Al$_x$Ni$_{1-x}$ exhibits a pronounced minimum at the stoichiometric composition of Al$_{0.50}$Ni$_{0.50}$. A similar observation for Al$_x$Sc$_{1-x}$ is depicted in Fig.~\ref{Fig:Resistivity}b. 

As discussed in Sec.~\ref{Sec:Resistivity_Intermet}, the increase in resistivity can be attributed to the generation of nonstoichiometric point defects, which introduce strong disorder scattering. The experimental findings indicate that compositional control at a level of better than 1 at.\%{} over the entire wafer is essential to achieve low and uniform resistivities, a demanding requirement for high-volume manufacturing. Furthermore, in both cases shown in Fig.~\ref{Fig:Resistivity}, low resistivities were only observed after high-temperature post-deposition annealing, likely due to thermally activated ordering (point defect reduction) and grain growth. Consequently, the compatibility of these annealing steps with the thermal budget of the device fabrication process must be carefully evaluated.

\begin{figure}[tb]
\includegraphics[width=12cm]{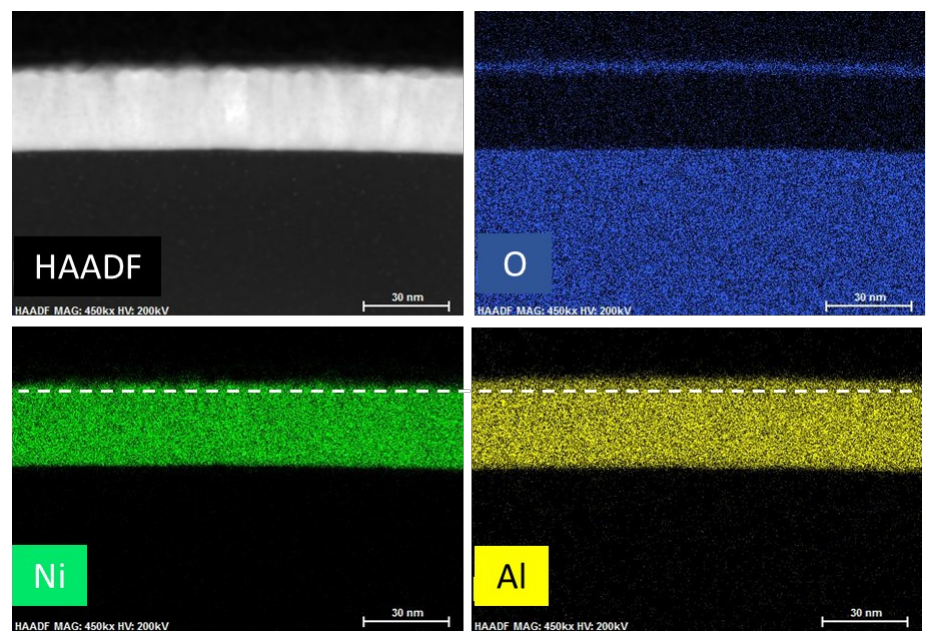} 
\caption{\label{Fig:NiAl EDS} High-angle annular dark field transmission electron micrograph as well as O, Al, and Ni energy-dispersive x-ray spectroscopy chemical maps of a 30 nm thick NiAl film (on SiO$_2$/Si) after air exposure. The chemical image analysis indicates the presence of an AlO$_x$ surface oxide.}
\end{figure}

As a further illustration of the challenges associated with binary intermetallics, Fig.~\ref{Fig:NiAl EDS} presents a cross-sectional transmission electron micrograph and the corresponding energy-dispersive x-ray spectroscopy chemical analysis of a NiAl film after air exposure. The chemical analysis reveals the presence of a native surface oxide, which deviates from the bulk stoichiometry, exhibiting a composition close to pure Al$_2$O$_3$.\cite{soulie_thickness_2020} This phenomenon can be attributed to element-specific surface processes, specifically metal outdiffusion governing the formation of the native oxide. This can significantly complicate compositional control, particularly for ultrathin films.\cite{cai_growth_2013} 

The tendency of forming nonstoichiometric native surface oxides has been observed in various aluminide intermetallics, although the specific oxide compositions may vary depending on the material system.\cite{adelmann_intermetallic_2023, soulie_al3sc_2024, soulie_AlCu_2024} Therefore, \textit{in situ} surface passivation techniques may be essential for the successful integration of (aluminide) intermetallics into scaled interconnects, both after deposition and potentially after patterning also.

\subsection{Ternary compounds \label{Sec:ternary}}

\begin{figure}[tb]
\includegraphics[width=16cm]{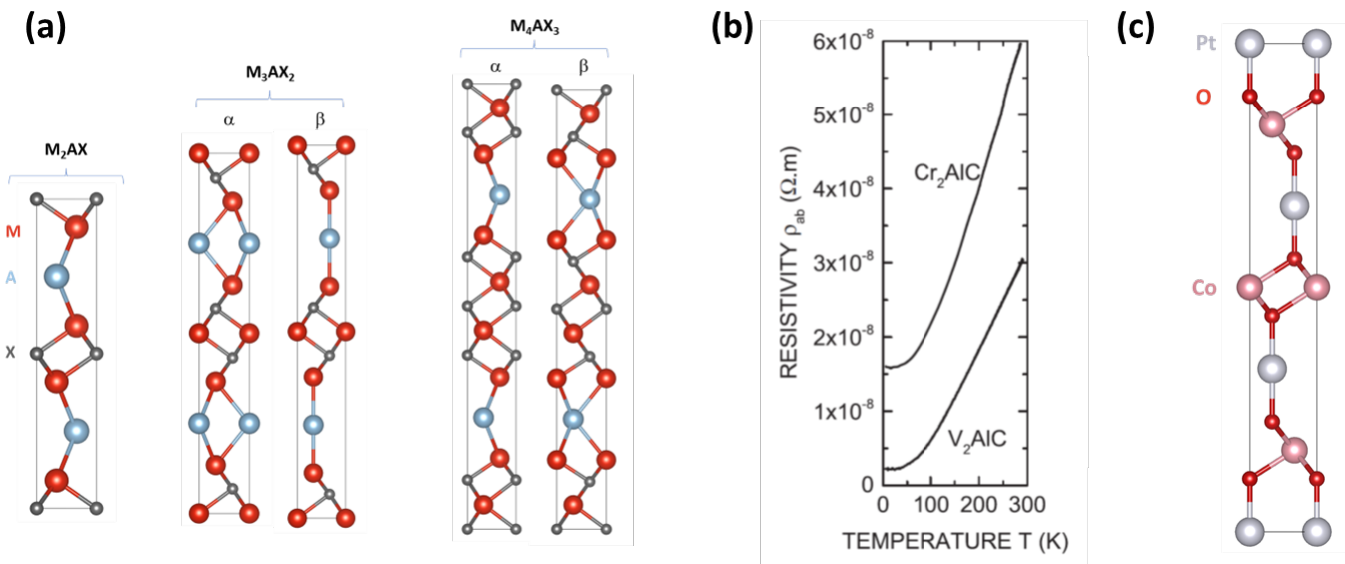} 
\caption{\label{Fig:MAX} (a) Crystal structure of MAX phases, M$_{n+1}$AX$_n$, where M is an early transition metal, A is an A-group element, X is C or N, and $n$ is an integer between 1 and 3. (b) Temperature-dependent in-plane resistivity of Cr$_2$AlC and V$_2$AlC single crystals. Reprinted with permission from Ref.~\onlinecite{ouisse_magnetotransport_2015}. (c) Crystal structure of the delafossite oxide PtCoO$_2$.}
\end{figure}

Beyond binary intermetallics, several ternary compounds have been explored for advanced interconnect metallization. Given the vast combinatorial space of ternary intermetallics and the limited knowledge of their properties, a comprehensive screening approach, akin to that employed for elemental or binary systems, is computationally prohibitive. Consequently, research has thus far focused on specific material classes, with particular attention given to MAX phases.\cite{adelmann_alternative_2018, moon_materials_2023} MAX phases are layered hexagonal carbide or nitride metallic ceramics, described by the generic formula M$_{n+1}$AX$_n$ (Fig.~\ref{Fig:MAX}a), where $1 \le n \le 3$; M is an early transition metal, A is an A-group element, and X is either C or N.\cite{barsoum_mn+1axn_2000,eklund_mn+1axn_2010,barsoum_max_2013,radovic_max_2013,gonzalez-julian_processing_2021,dahlqvist_max_2024} 

MAX phases typically exhibit significant thermal and electrical conductivity, along with high melting points. Certain MAX compounds, such as Cr$_2$AlC and V$_2$AlC, demonstrate bulk in-plane resistivities on the order of 10 $\mu\Omega$cm at room temperature (Fig.~\ref{Fig:MAX}b).\cite{ouisse_magnetotransport_2015} An \textit{ab initio} screening study has identified low $\rho_0 \times \lambda$ products (see Sec.~\ref{Subsec:Rholambda}) for several MAX phases, confirming their potential for scaled interconnect metallization.\cite{sankaran_ab_2021}

Another material system of potential interest are metallic delafossite oxides, particularly PdCoO$_2$ and PtCoO$_2$.\cite{harada_thin-film_2021} These layered hexagonal compounds (see Fig.~\ref{Fig:MAX}c) exhibit ultralow bulk resistivities comparable to that of aluminum,\cite{mackenzie_properties_2017, harada_thin-film_2021,harada_metallic_2022,zhang_crystal-chemical_2024} along with exceptionally long mean free paths.\cite{bachmann_directional_2022} Recent experimental studies have reported thin films with thicknesses within the target range for interconnect applications.\cite{brahlek_growth_2019,ok_pulsed-laser_2020,hagen_atomic_2022,harada_sputter-grown_2023} However, further experimental investigation is required to thoroughly assess the scalability and suitability of these delafossite oxides for interconnect line integration.

Ternary compounds, like their binary counterparts, are likely to encounter similar challenges in terms of composition control and processing. Furthermore, both MAX phases and delafossite compounds are highly anisotropic conductors, exhibiting low resistivity in the in-plane directions. While this anisotropy could be advantageous by suppressing surface scattering at top surfaces or interfaces (see Sec.~\ref{Sec:AnisoModel}),\cite{de_clercq_resistivity_2018, kumar_ultralow_2022} achieving the desired crystallographic orientation becomes critical. Specifically, fully (001)-textured films without misoriented grains must be realized to harness these properties. Therefore, extensive experimental validation is still required to confirm the viability of these ternary metals for interconnect applications.

\subsection{Beyond binary and ternary intermetallics: one-dimensional metals and topological Weyl semimetals}

As discussed in Sec.~\ref{Sec:AnisoModel}, anisotropic resistivity and reduced dimensionality can mitigate surface scattering. One-dimensional conductors, in particular, have been proposed as ideal interconnect materials due to their ability to suppress surface scattering at the top, bottom, and sidewalls of interconnect lines.\cite{kumar_ultralow_2022} Unlike two-dimensional layered metals (MAX, delafossite oxides), which exhibit low resistivity in two directions but higher resistivity in the perpendicular direction, one-dimensional metals possess a single direction of low resistivity with significantly increased resistivity in the two orthogonal directions.

The suppression of surface scattering in one-dimensional metals can be incorporated into both resistivity simulations\cite{VanTroeye2023} and material benchmarking (see Sec.~\ref{Sec:TF_NW}).\cite{kumar_ultralow_2022} Similar to the $\rho_0\times\lambda$ tensor introduced in Sec.~\ref{Subsec:Rholambda}, a figure of merit for nanowires has been defined in Eq.~\eqref{eq:rholambdau_bvt} that accounts for the reduction in surface scattering.\cite{VanTroeye2023, kumar_ultralow_2022} Potential one-dimensional metal candidates include binary hexagonal intermetallics (\textit{e.g.}, CoSn, OsRu), orthorhombic intermetallics (\textit{e.g.}, VPt$_2$, MoNi$_2$), and ternary borides (\textit{e.g.}, YCo$_3$B$_2$). 

However, it should be noted that no thin film results have yet unambiguously demonstrated the suppression of surface scattering in these materials. Furthermore, integrating such materials into interconnects will necessitate single crystals with the low resistivity axis aligned with the interconnect wires. Currently, no viable manufacturing pathways exist for producing such interconnects, indicating that significant research and development are still needed to realize the potential of these materials.

Additionally, topological semimetals, which encompass both Weyl and multifold-fermion semimetals, have recently emerged as promising candidates for future interconnect technologies. Weyl semimetals are distinguished by their unique electronic structure, characterized by linear band dispersion, degenerate Weyl nodes, and topologically protected surface states.\cite{Xu_2015_Weyl} These surface states exhibit high conductivity and are robust against disorder. Examples of Weyl semimetals include TaAs,\cite{lv_2015_observation} TaP, NbAs,\cite{zhang_2019_NbAs} MoP,\cite{Han_2023_MoP} and NbP, while CoSi,\cite{Chen_2020_IEDM, lien_2023_CoSi} RhSi, and AlPt represent multifold-fermion semimetals. 

In the case of CoSi, calculations have shown that the effective resistivity (resistance normalized by cross-sectional area) decreases as wire dimensions are reduced, even in the presence of grain boundaries, due to the dominance of surface-driven transport channels.\cite{Lanzillo_2022} Experimental evidence for the topological semimetal NbAs suggests that indeed the electrical resistivity can decrease as the cross-sectional area decreases,\cite{zhang_2019_NbAs} although further investigations are required to confirm these findings in interconnect-relevant geometries. Similar to one-dimensional metals, topological semimetals will require the fabrication of single-crystal (epitaxial) wires. Furthermore, the reliability of interconnects based on topological semimetals remains uncertain, necessitating further fundamental research to evaluate their viability for scalable interconnect applications.

\section{Resistance Modeling for nanoscale interconnect lines
\label{Sec:Interconnect_Model}}

The resistivity trends presented in Fig.~\ref{fig:ms5} can be leveraged to develop calibrated models for interconnect line resistance, enabling benchmarking against (projected) Cu line resistance values at scaled dimensions. A simplified geometrical model for Ru and Ir interconnects, defined by a width $w$, height $h = w \times \mathrm{AR}$, and aspect ratio (AR), has recently been introduced based on the data from Fig.~\ref{fig:ms5}.\cite{adelmann_alternative_2018} To project barrierless line resistance as a function of line width $w$, resistivity trends as a function of cross-sectional area were derived from the data for Ru (after post-deposition annealing at 420\deg C). The Ru metallization scheme also incorporated a 0.3 nm thick adhesion liner.\cite{wen_atomic_2016} For comparison, Cu resistivities were taken from an established line resistance model.\cite{ciofi_impact_2016}

\begin{figure}[tb]
\includegraphics[width=8.5cm]{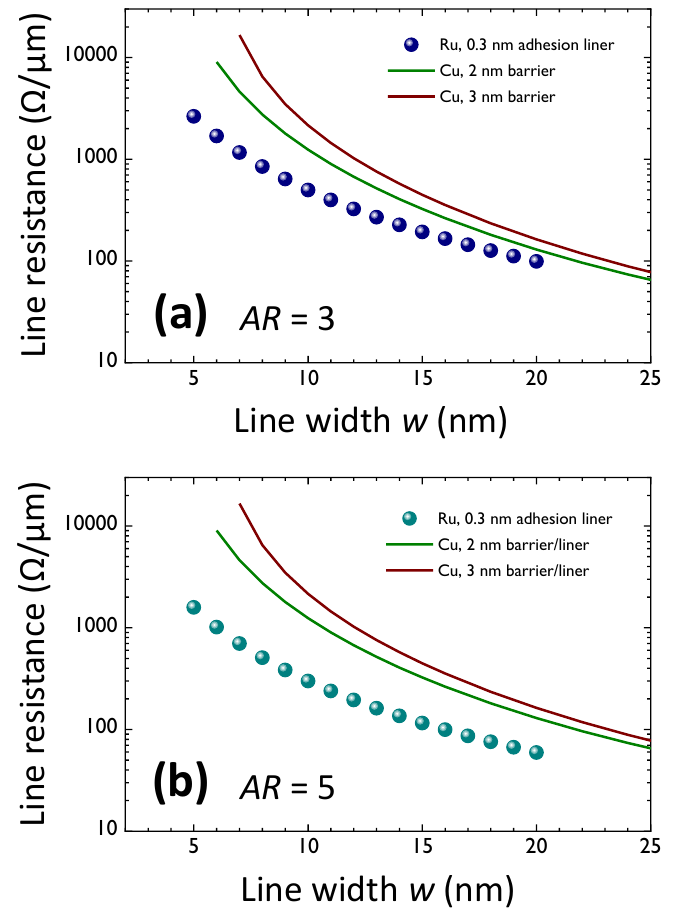} 
\caption{\label{Fig:Line_Resistance_model} Projected line resistance for Ru and Cu interconnects with aspect ratios of (a) 3 and (b) 5, respectively. The Ru model includes a 0.3 nm adhesion liner, while the Cu trend lines incorporates barrier and liner layers with aggregate  thicknesses of 2 and 3 nm. The data is based on calibrated models for Ru\cite{adelmann_alternative_2018} and Cu.\cite{ciofi_impact_2016}}
\end{figure}

The projected line resistances for Ru and Cu metallization, considering different combined diffusion barrier and liner thicknesses, are shown in Fig.~\ref{Fig:Line_Resistance_model} for ARs of 3 (Fig.~\ref{Fig:Line_Resistance_model}a) and 5 (Fig.~\ref{Fig:Line_Resistance_model}b). The data demonstrate a significant potential for lower line resistances with Ru interconnects compared to Cu, even with scaled barrier and liner layers, particularly at higher aspect ratios. For instance, the model suggests that Ru interconnect lines could achieve a threefold reduction in line resistance over Cu when the total barrier and liner thickness is 2 nm, with a line width of $w = 8$ nm and an AR of 3.

Moreover, these resistance models can provide insights into the mechanisms driving the crossover in line resistance between Cu- and Ru-based interconnects. A quantitative comparison of Ru and Cu resistivities (see Fig.~\ref{fig:ms5}) indicates that the scaling advantage of Ru does not arise from a lower resistivity but from the increased conductor volume when the thicker barrier and liner layers required for Cu are replaced by a much thinner adhesion layer. Nevertheless, resistivity scaling remains a crucial factor as high resistivities at low dimensions due to poor resistivity scaling can negate the potential benefits of increased conductor volume. Therefore, barrierless metallization and favorable resistivity scaling must be complement each other to realize low line resistances in scaled interconnects.

\begin{figure}[tb]
\includegraphics[width=16cm]{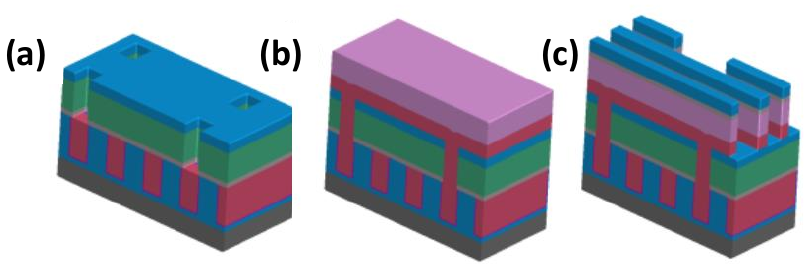} 
\caption{\label{Fig:Semidamascene_flow} Schematic representation of the semidamascene interconnect integration process: (a) via patterning in low-$\kappa$ dielectric (green) using a hardmask (blue); (b) metal filling of the etched vias (red-brown) followed by metal overfill (light purple); (c) line patterning in the overfilled metal layer (light purple) using a second hardmask (blue).\cite{murdoch_SDamasc_2020} }
\end{figure}

A second factor contributing to reduced line resistance is an increased AR.\cite{tokei_inflection_2020} While AR is not an intrinsic material property, it is strongly influenced by the integration process. Conventional dual-damascene Cu interconnects (Fig.~\ref{DD_flow}) are limited to ARs of 2 to 3 due to the challenges of Cu filling. However, alternative process schemes, such as semidamascene integration (Fig.~\ref{Fig:Semidamascene_flow}), can enable higher ARs by combining damascene via filling with direct metal etching. While Cu etching of scaled high-AR lines remains difficult, Ru and Mo are much better suited for reactive-ion etching (see Sec.~\ref{Sec:ProcessTech} and Tab.~\ref{Tab:Process}). Specifically, Ru lines with aspect ratios up to 6 and metal pitches as small as 18 nm have been demonstrated.\cite{decoster_patterning_2022} As shown in Fig.~\ref{Fig:Line_Resistance_model}b, increasing the aspect ratio to 5 can result in a fivefold reduction in line resistance for Ru compared to Cu (at an aspect ratio of 2) with a 2 nm barrier/liner thickness and a line width of 8 nm. Therefore, the combination of alternative metals with novel integration schemes offers significant potential for reducing line resistance in scaled interconnects.

\section{Material Considerations for Interconnect Processing}

Integrating alternative metals into scaled interconnects often requires the development of novel unit processes and metallization modules within the final stages of the alternative metal workflow depicted in Fig.~\ref{fig:workflow}. While a comprehensive review of available process technologies, their maturity, and their limitations is beyond the scope of this tutorial, we will introduce several topics that become increasingly critical in the development of scaled interconnect line manufacturing. This section will conclude with a brief evaluation of the current maturity of key unit processes for selected alternative metals.   

\subsection{Adhesion and stress}

A fundamental property of interconnect metallization is its adhesion to surrounding low-$\kappa$ dielectrics. Metal--dielectric interfaces often exhibit weaker adhesion compared to metal--metal or dielectric--dielectric interfaces, potentially leading to metal film delamination and catastrophic failure. While deposition conditions impact adhesion, it can be considered as a material-dependent property. Noble metals generally exhibit weaker adhesion to dielectrics than base metals due to weaker interfacial bonding. High-quality graphene also suffers from poor adhesion due to weak van der Waals interactions with surrounding dielectrics or metals.

Adhesion can be enhanced by incorporating adhesion liners between the main metal (\textit{e.g.}, a noble metal) and the surrounding dielectrics. However, like diffusion barriers, adhesion liners occupy interconnect volume and typically contribute minimally to conductance. Therefore, minimizing their thickness is crucial. Experimental studies have demonstrated that base metals such as Mo exhibit strong adhesion to common dielectrics.\cite{founta_properties_2022} This allows for Mo metallization without the need for additional barriers or liners. 

In contrast, the more noble metal Ru requires an adhesion liner (\textit{e.g.}, TiN or TaN) due to its weaker adhesion to dielectrics. Nevertheless, studies have shown that the liner thickness can be reduced to as little as 0.3 nm without compromising its effectiveness.\cite{wen_atomic_2016} This suggests that even non-continuous films can function as adhesion liners and may be more scalable than diffusion barriers. Even more noble metals like Ir and Rh however require further investigation to ensure adequate adhesion and prevent delamination.

Delamination can be exacerbated by high built-in stress within the metallization stack, further weakening the interface between metals and dielectrics. Additionally, the combination of high compressive stress and capillary forces during filling can lead to nanostructure deformation, such as line wiggling.\cite{motoyama_metal-induced_2022} Stress is not an intrinsic material property but is mainly determined by the deposition method. Physical vapor deposited films often exhibit high (tensile) stress after deposition, which is typically generated during island coalescence at the initial stages of nucleation and growth. However, the overall stress behavior can be complex.\cite{messier_revised_1984,windischmann_intrinsic_1992,thompson_structure_2000,petrov_microstructural_2003,bhandari_competition_2007,abadias_stress_2018} 

For instance, as-deposited PVD Mo films have been observed to have built-in tensile stress as high as 1500 MPa, depending on the film thickness.\cite{founta_properties_2022} Post-deposition annealing and associated grain growth can significantly modify stress, even leading to compressive stress after cooling within certain temperature ranges.\cite{founta_properties_2022,founta_stress_2022} While stress management is primarily a topic for deposition process development, it is particularly critical for metals with inherently weak adhesion.

\subsection{Oxidation resistance}

During interconnect patterning, certain metal surfaces may be exposed to air or other reactive environments, making chemical inertness, particularly oxidation resistance, a critical consideration. Even self-limiting surface oxidation processes can result in the formation of native oxide layers, typically around 2 nm thick, consuming approximately 1 nm of metal.\cite{founta_properties_2022} For scaled metal lines with dimensions on the order of 10 nm, surface oxidation must therefore be strictly avoided. Noble metals are inherently more chemically inert than base metals and thus offer greater resistance to oxidation. While this also leads to weaker adhesion, it renders noble metals more compatible with interconnect process flows.

For compound metals, the situation is even more complex. As discussed in Sec.~\ref{Sec:binary} for NiAl (see Fig.~\ref{Fig:NiAl EDS}), surface oxides of binary metals can be nonstoichiometric, leading to compositional changes in the region immediately beneath the surface oxide.\cite{soulie_thickness_2020,soulie_al3sc_2024} In such material systems, controlling the composition of scaled interconnect lines is extremely challenging, and surface oxidation must thus be strictly avoided. While \textit{in situ} patterning and passivation or capping can potentially address these issues, they introduce significant process complexity and should be carefully considered during metal selection.

\subsection{Process technology readiness\label{Sec:ProcessTech}}

As mentioned above, a comprehensive review of the state-of-the-art process technology for Cu and alternative metal integration is beyond the scope of this tutorial. However, from a general perspective, both dual-damascene (Fig.~\ref{DD_flow}) and semidamascene integration routes (Fig.~\ref{Fig:Semidamascene_flow}) require critical unit processes for manufacturing scaled interconnects. These include metal trench and via filling, typically accomplished through electroplating or chemical vapor deposition. Line definition requires chemical-mechanical polishing (CMP) for dual-damascene and reactive-ion etching (RIE) for semidamascene integration. While Cu is well-suited for dual-damascene integration due to the availability of mature chemical-mechanical polishing processes, its suitability for semidamascene integration is limited by the challenges associated with Cu reactive-ion etching.

\begin{figure}[tb]
\includegraphics[width=7.5cm]{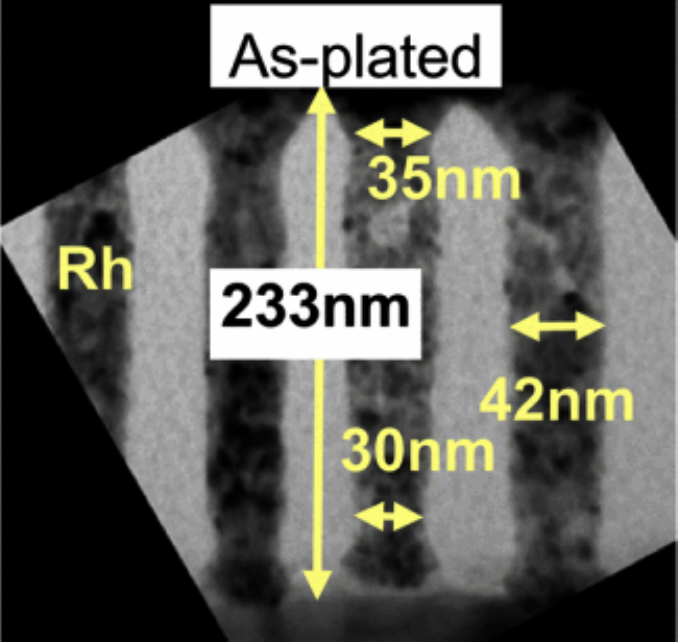} 
\caption{\label{Fig:Plated_Rh} Cross-sectional transmission electron micrograph of a Rh-filled interconnect line with a width of $< 40$ nm. Reprinted with permission from Ref.~\onlinecite{shao_alternative_2007}.}
\end{figure}

For alternative metals, the availability of suitable chemical-mechanical polishing and/or reactive-ion etching processes, in addition to deposition techniques, is thus crucial. Ru has demonstrated excellent results with reactive-ion etching, enabling scaled lines with high aspect ratios and precise sidewall control.\cite{wan_subtractive_2018,murdoch_SDamasc_2020,decoster_patterning_2022} However, for many other alternative metals, both chemical-mechanical polishing and reactive-ion etching remain underdeveloped. Rh, for example, offers low resistivity, a high melting point, and potential for high electromigration resistance.\cite{lanzillo_reliability_2022} Adhesion engineering remains however a challenge, particularly when minimizing the thickness of adhesion liners to avoid reducing conductor volumes. Rh can be electroplated\cite{son_kinetics_2010}, and the filling of sub-40 nm wide lines and vias with high aspect ratios has been demonstrated (Fig.~\ref{Fig:Plated_Rh}).\cite{shao_alternative_2007} Yet, dual-damascene integration requires chemical-mechanical polishing, which is not well-established for Rh and requires aggressive abrasives and oxidizers.\cite{shao_alternative_2007} Moreover, the lack of manufacturable reactive-ion etching processes for Rh hinders semidamascene integration, making it a significant obstacle to realizing the potential of Rh metallization in high-volume manufacturing CMOS circuits.

Table~\ref{Tab:Process} provides a summary of the process readiness for selected elemental metals, as well as binary and ternary compounds. The table provides the authors' assessment of the process maturity for each material as of the time of publication: a `+' indicates a mature process suitable for high-volume manufacturing, an `o' signifies the availability of a risk process, and a `-' denotes a process that is currently at the research stage. The table highlights the intricate and lengthy journey required for the successful integration of alternative metals into manufacturable interconnects. Mo and Ru processes appear to be the most mature, making them leading candidates for high-volume manufacturing, as already outlined in Sec.~\ref{Sec:Downselection}. In contrast, many critical process steps are still lacking for other metals. 

\begin{table}
\caption{\label{Tab:Process}Summary of process readiness for high-volume manufacturing of selected alternative metals at time of publication: compatibility of metal process temperatures with logic back-end-of-line (BEOL) thermal budget, maturity of chemical-mechanical polishing, reactive-ion etching, as well as wet cleaning unit processes, as well as relative extendability of alternative metallization to future technology nodes.}\textcolor{ForestGreen}{\Large +} : mature process; \textcolor{BurntOrange}{\large o} : risk process; \textcolor{Red}{\Large -} : research process.
    \centering
    \begin{tabular}{lcccccc}
\toprule
Metal & Logic BEOL & Trench filling & Chemical-mechanical & Reactive ion & Wet cleaning & Extendability \\
& thermal budget & & polishing & polishing & \\
\hline
Cu & \textcolor{ForestGreen}{\Large +} & \textcolor{ForestGreen}{\Large +} & \textcolor{ForestGreen}{\Large +} & \textcolor{Red}{\Large -} & \textcolor{ForestGreen}{\Large +} & \textcolor{Red}{\Large -}\\
W & \textcolor{ForestGreen}{\Large +} & \textcolor{ForestGreen}{\Large +} & \textcolor{ForestGreen}{\Large +} & \textcolor{ForestGreen}{\Large +} & \textcolor{ForestGreen}{\Large +} & \textcolor{Red}{\Large -}\\
Mo & \textcolor{BurntOrange}{\large o} & \textcolor{BurntOrange}{\large o}& \textcolor{BurntOrange}{\large o} & \textcolor{BurntOrange}{\large o} & \textcolor{BurntOrange}{\large o} & \textcolor{ForestGreen}{\Large +}\\
Ru & \textcolor{ForestGreen}{\Large +} & \textcolor{BurntOrange}{\large o} & \textcolor{BurntOrange}{\large o} & \textcolor{ForestGreen}{\Large +} & \textcolor{BurntOrange}{\large o} & \textcolor{ForestGreen}{\Large +}\\
Ir & \textcolor{ForestGreen}{\Large +} & \textcolor{Red}{\Large -} & \textcolor{Red}{\Large -} & \textcolor{Red}{\Large -} & \textcolor{Red}{\Large -} & \textcolor{BurntOrange}{\large o}\\
Rh & \textcolor{ForestGreen}{\Large +} & \textcolor{Red}{\Large -} & \textcolor{Red}{\Large -} & \textcolor{Red}{\Large -} & \textcolor{Red}{\Large -} & \textcolor{BurntOrange}{\large o}\\
NiAl & \textcolor{ForestGreen}{\Large +} & \textcolor{Red}{\Large -} & \textcolor{Red}{\Large -} & \textcolor{Red}{\Large -} & \textcolor{Red}{\Large -} & \textcolor{BurntOrange}{\large o}\\
CuAl$_x$ & \textcolor{ForestGreen}{\Large +} & \textcolor{Red}{\Large -} & \textcolor{Red}{\Large -} & \textcolor{Red}{\Large -} & \textcolor{Red}{\Large -} & \textcolor{BurntOrange}{\large o}\\
Al$_3$Sc & \textcolor{ForestGreen}{\Large +} & \textcolor{Red}{\Large -} & \textcolor{Red}{\Large -} & \textcolor{Red}{\Large -} & \textcolor{Red}{\Large -} & \textcolor{BurntOrange}{\large o}\\
PtCoO$_2$ & \multirow{2}{*}{\textcolor{Red}{\Large -}} & \multirow{2}{*}{\textcolor{Red}{\Large -}} & \multirow{2}{*}{\textcolor{Red}{\Large -}} & \multirow{2}{*}{\textcolor{Red}{\Large -}} & \multirow{2}{*}{\textcolor{Red}{\Large -}} & \multirow{2}{*}{\textcolor{BurntOrange}{\large o}} \\
(delafossite) & \\
Cr$_2$AlC & \multirow{2}{*}{\textcolor{Red}{\Large -}} & \multirow{2}{*}{\textcolor{Red}{\Large -}} & \multirow{2}{*}{\textcolor{Red}{\Large -}} & \multirow{2}{*}{\textcolor{Red}{\Large -}} & \multirow{2}{*}{\textcolor{Red}{\Large -}} & \multirow{2}{*}{\textcolor{BurntOrange}{\large o}} \\
(MAX) & \\
\toprule
    \end{tabular}
\end{table}

\section{Sustainability of alternative metals}

Traditionally, the selection of alternative interconnect metals has mainly considered technological, physical, and economic factors. However, recognizing the increasing importance of sustainability, this section introduces a streamlined framework for assessing the sustainability of alternative interconnect metals. This framework incorporates seven sustainability aspects (SAs) and evaluates examples of selected current and emerging interconnect metals (Cu, Al, Ni, Ru, Co, Mo, Ir, Rh). To avoid shifting environmental burdens, a life cycle approach is essential, requiring consideration of the integration method for alternative interconnect metals. Understanding material and energy flows during integration is crucial for assessing overall sustainability. While processes with fewer steps may reduce environmental impact, energy requirements during integration must also be carefully considered. This section discusses these integration considerations and provides a condensed overview of the sustainability assessment framework detailed in Ref.~\onlinecite{boakes2024selection}.

The proposed sustainability assessment framework for alternative interconnect metals is categorized into seven sustainability aspects (SAs), each with at least one indicator to quantify its impact. SA1 focuses on supply risk, using the Herfindahl-Hirschman index (HHI) to assess market concentration.\cite{bromberg_herfindahl-hirschman_nodate,noauthor_antitrust_2015} The Herfindahl-Hirschman index values in Tab.~\ref{Tab:Sustain} were extracted from Refs.~\onlinecite{world_mining_data_2023} and \onlinecite{raw_materials_profiles}. SA2 addresses criticality and conflict, considering metals listed as critical raw materials (CRMs) in the US\cite{Fortier2022} and EU\cite{Commission2023}, as well as those on the EU conflict mineral list \cite{EUConflictmaterials}. SA3 evaluates metal circularity within integrated circuit manufacturing processes, and acknowledges the challenges of calculating the site material circularity index (CI) \cite{ANSI/UL}. SA4 assesses climate change impact through global warming potential (GWP) values \cite{Nuss2014}, while SA5 focuses on water scarcity using the EF~3.1 methodology\cite{noauthor_developer_nodate} for upstream water use. SA6 examines the impact on natural resources through abiotic resource depletion potential (ADP) values\cite{van_oers_abiotic_2020}, and SA7 assesses human health impacts using EF~3.1 methodologies such as ``human toxicity cancer and non-cancer'' and ``particulate matter''. These indicators collectively provide a comprehensive evaluation of the sustainability of alternative interconnect metals. For more detailed information on each SA and its associated indicator(s), please refer to Ref.~\onlinecite{boakes2024selection}.

\begin{sidewaystable}
\centering
\caption{\label{Tab:Sustain}Summary table of sustainability aspect (SA) indicators for current and alternative interconnect metals. The volumetric impact values for SA4 to SA7 have been quantified based on the cradle-to-gate production of 1 cm$^3$ of interconnect metal. The values for such indicator have been classified as green, amber or red as defined in Ref.~\onlinecite{boakes2024selection}.}
\begin{tabular}{ccccccccc}
\toprule
& Density \cite{density} & SA1: HHI \cite{world_mining_data_2023, raw_materials_profiles}& SA2 & SA4 & SA5: WS \cite{boakes2024selection} & SA6: ADP \cite{van_oers_abiotic_2020} & SA7& SA7\\
& & & (Refs.~\onlinecite{Commission2023, Fortier2022, EUConflictmaterials}) & Embedded GWP \cite{Nuss2014} & & & Human Toxicity \cite{Nuss2014}& Particulates \cite{boakes2024selection}\\	
& {kg/m$^3$} & (0--10000) & & (kg CO$_2$/cm$^3$) & (m$^3$/cm$^3$) &	(kg Sb eq/cm$^3$ & (CTUh/cm$^3$) &	(Disease incidences/cm$^3$) \\ \hline
Cu & 9.0& \textcolor{ForestGreen}{1097}& \textcolor{BurntOrange}{Yes/No/No} & \textcolor{ForestGreen}{0.0251} & \textcolor{ForestGreen}{0.02} & \textcolor{ForestGreen}{$2.42\times 10^{-04}$} & \textcolor{ForestGreen}{$2.42\times 10^{-06}$} & \textcolor{ForestGreen}{$5.11\times 10^{-09}$}\\
Ni & 8.9 & \textcolor{BurntOrange}{2110} & \textcolor{BurntOrange}{No/Yes/No} & \textcolor{ForestGreen}{0.0579} & \textcolor{ForestGreen}{0.03} & \textcolor{ForestGreen}{$1.07\times 10^{-05}$} & \textcolor{ForestGreen}{$2.05\times 10^{-07}$} & \textcolor{ForestGreen}{$8.27\times 10^{-08}$}\\
Mo & 10.0 & \textcolor{BurntOrange}{2,266} & \textcolor{ForestGreen}{No/No/No} & \textcolor{ForestGreen}{0.0583} & \textcolor{ForestGreen}{0.02} & \textcolor{BurntOrange}{$2.25\times 10^{-03}$} & \textcolor{BurntOrange}{$9.20\times 10^{-06}$} &	\textcolor{ForestGreen}{$2.00\times 10^{-09}$}\\
Al & 2.7 & \textcolor{red}{3372} &  \textcolor{red}{Yes/Yes/No} & \textcolor{ForestGreen}{0.0222} & \textcolor{ForestGreen}{0.01} & \textcolor{ForestGreen}{$1.13\times 10^{-10}$} & \textcolor{ForestGreen}{$1.46\times 10^{-08}$} & \textcolor{ForestGreen}{$6.84\times 10^{-09}$}\\
Co & 9.0 & \textcolor{red}{4,876} & \textcolor{red}{Yes/Yes/No} & \textcolor{ForestGreen}{0.0739} & \textcolor{ForestGreen}{0.34} & \textcolor{ForestGreen}{$4.18\times 10^{-06}$} & \textcolor{ForestGreen}{$3.38\times 10^{-08}$} & \textcolor{ForestGreen}{$3.12\times 10^{-08}$}\\
Ru & 12.4 & \textcolor{red}{8,718} & \textcolor{red}{Yes/Yes/No} & \textcolor{BurntOrange}{26} & \textcolor{red}{164} & \textcolor{red}{$3.34$} & \textcolor{BurntOrange}{$1.98E\times 10^{4}$} &	---\\
Rh & 12.4 & \textcolor{red}{7352} & \textcolor{red}{Yes/Yes/No} & \textcolor{red}{436} & \textcolor{BurntOrange}{152} & \textcolor{ForestGreen}{$2.61\times 10^{-05}$} & \textcolor{red}{$3.35\times 10^{-03}$} & \textcolor{red}{$4.75\times 10^{-05}$}\\
Ir & 22.4 & \textcolor{red}{7986} & \textcolor{red}{Yes/Yes/No} & \textcolor{red}{198} & \textcolor{red}{200} &	\textcolor{BurntOrange}{3.14} & \textcolor{red}{$1.12E\times 10^{-03}$} &	---\\
\toprule
\end{tabular}
\end{sidewaystable}

Table~\ref{Tab:Sustain} provides a summary of the sustainability performance of the examined interconnect metals, presenting a nuanced perspective on the seven proposed indicators. The sustainability impacts (SA4 to SA7 in Tab.~\ref{Tab:Sustain}) are calculated based on the cradle-to-gate impact to produce 1 cm$^3$ of metal. This calculation assumes that the volume of the final deposited layer of interconnect metal is independent of the metal. However, the volume ratio required to achieve the desired final volume deposited should be considered. 

Evaluating metal deposition efficiency $\eta_\mathrm{dep}$ is essential to determine the actual used volume $V_\mathrm{u}$. Typical deposition processes exhibit a range of $\eta_\mathrm{dep}$ values spanning from 1 to 20\% for chemical-vapor-based deposition processes\cite{weber_assessing_2023} but can be much higher for physical vapor processes. Additionally, subtractive integration schemes lead to further material loss determined by the material use efficiency of the integration process $\eta_\mathrm{int}$, and influenced by the choice of interconnect metal. By contrast, damascene integration schemes require the depositions of large overburden before chemical-mechanical polishing, leading also to $\eta_\mathrm{int}\ll 1$.  $V_\mathrm{u}$ can be defined as

\begin{equation}
\label{E:Use_Volume}
V_\mathrm{u} = \frac{V_\mathrm{IC}}{\eta_\mathrm{dep}\eta_\mathrm{int}},    
\end{equation}

\noindent where $V_\mathrm{IC}$ is the volume of the manufactured interconnect (layer), determined by interconnect dimensions as well as circuit-specific via and line densities. 

A more accurate assessment of the environmental impact of the interconnect metal a for specific sustainability aspect, $IX_\mathrm{met}$, can be obtained by

\begin{equation}
\label{E:SA_vol}
IX_\mathrm{met} = V_\mathrm{u} \times \mathrm{SA}X_\mathrm{vol},
\end{equation}

\noindent with the volumetric impacts for SA4 to SA7 in Tab.~\ref{Tab:Sustain}.

The preceding introduction outlines a streamlined sustainability assessment methodology for alternative interconnect metals.\cite{boakes2024selection} The proposed seven sustainability indicators offer a holistic, life cycle perspective, enabling a comprehensive evaluation of sustainability. A qualitative analysis of the volumetric impact values in Tab.~\ref{Tab:Sustain} can aid process engineers in identifying trade-offs and making informed decisions for developing advanced interconnect applications. Notably, Al, Ni, Co, and Mo exhibit relatively favorable performance in at least three of the seven indicators, while the platinum group metals (Ru, Ir, and Rh) demonstrate comparatively poor results in at least six of the seven indicators.

Further analysis involves multiplying the volumetric impact in Tab.~\ref{Tab:Sustain} by the total volume of metal consumed to achieve a fixed function, \textit{i.e.}, a set volume of deposited metal. This incorporates the material use efficiency inherent in the deposition and integration methods. Moreover, the application of normalization or weighting factors is recommended to prioritize sustainability indicators based on the situational circumstances such as company specific sustainability goals, willingness to take financial risks, or location specific regulations/accessibility to materials. Combined with the technological assessment, this streamlined sustainability methodology offers decision makers a foundation for expanding criteria in the selection of alternative metals for advanced interconnect applications.

\section{Summary and Conclusions}

The scaling of the interconnect metal pitch is today a crucial challenge in the development of advanced microelectronic technology. As the transistor pitch approaches its physical limits, reducing metal wire pitch has become a primary strategy for further shrinking circuit area. While transistor stacking can still increase density, it also necessitates tighter metal pitches to prevent interconnect congestion, potentially offsetting the benefits of stacking. Furthermore, the interconnect $RC$ delay poses a significant constraint on the throughput of CMOS circuits, even at current technology nodes. To keep $RC$ under control, both the resistance ($R$) and capacitance ($C$) of the interconnect must be optimized. Optimizing $R$ involves maximizing the metal fill factor within lines and vias while using a metal with the lowest possible resistivity. Optimizing $C$ requires the use of low-$\kappa$ dielectrics or incorporating air gaps between lines, which is beyond the scope of this tutorial.

The current Cu-based dual-damascene metallization scheme, introduced in 1999, is facing increasing challenges. To ensure reliability, Cu metallization requires barrier layers to prevent Cu diffusion into the surrounding dielectrics, which can cause dielectric breakdown. Today, TaN has emerged as the standard barrier material. Electromigration criteria further necessitate the inclusion of Co liner layers between Cu and TaN, as well as on the top of the Cu line (Co all-around liners). Both TaN and Co layers occupy a substantial portion of scaled interconnects while contributing minimally to line conductance. However, reducing the thickness of these layers without compromising their functionality is increasingly difficult. Despite ongoing efforts, achieving a combined thickness of 1 nm remains challenging. Even at a line width of 10 nm, a combined barrier and liner thickness of 1 to 1.5 nm would occupy 20 to 30\%{} of the line volume, significantly impacting line resistance. Moreover, the increasing difficulty of void-free filling of narrow lines using the damascene process suggests that Cu dual-damascene metallization may become unsustainable for metal pitches below 20 nm.

In the last decade, the limitations of Cu-based dual-damascene metallization have spurred the search for alternative metals. Given the relatively simple structure of interconnects, advancements in this field are mainly driven by material choices, making alternative interconnect metallization an exciting area of materials science. As demonstrated in this tutorial, the pursuit of novel interconnect metals requires a multifaceted approach that considers various criteria. While line resistance is paramount, reliability and thermal aspects must not be overlooked. As illustrated by the calibrated narrow line models in Sec.~\ref{Sec:Interconnect_Model}, achieving low line resistance necessitates a conductor metal with low resistivity and barrierless metallization to maximize conductor volume. Therefore, promising alternative metals must meet dielectric breakdown and electromigration criteria without the need for barriers. Other important material properties include adhesion to surrounding dielectrics, built-in stress, and oxidation resistance. Furthermore, process readiness and sustainability considerations should not be neglected.

The combination of resistance and reliability criteria has led to a focus on refractory metals---which promise high barrierless reliability---with a short mean free path for charge carriers, low bulk resistivity, and thus with low resistivity at nanoscale dimensions. While research initially centered on elemental metals, it has more recently expanded to include binary and ternary intermetallics. Among the materials studied, Ru and Mo emerge as the most promising candidates. Currently, the semiconductor industry is investing significant resources in developing the necessary process technology to integrate these metals into sub-10 nm interconnect lines without barriers. Their favorable etch characteristics also enable alternative integration routes, such as semidamascene integration, which can potentially facilitate higher aspect ratio lines, further reducing line resistance. Consequently, Ru and Mo are expected to be integrated into logic and memory devices in future technology nodes within the next decade.

Additional promising conductor materials include intercalated graphene and, in the longer term, topological materials such as Weyl semimetals. While these materials, including binary and ternary intermetallics, are currently being studied as thin films, their behavior in scaled wires remains to be explored. As discussed in Sec.~\ref{Sec:binary}, integrating these materials into interconnects presents significantly greater challenges compared to elemental metals. The development of manufacturable process technology for these materials is still in its early stages. However, the growing interest in such a diverse range of materials, with the potential for further discoveries, suggests that this field will remain a dynamic and exciting area of materials science for years to come.

\begin{acknowledgments}

The authors would like to thank Shibesh Dutta (ASM Netherlands), Anshul Gupta (imec), Kristof Moors (FZ J\"ulich, imec), Valeria Founta (KU Leuven, imec), Nancy Heylen (imec), Johan Meersschaut (imec), Jeroen Scheerder (imec), Olivier Richard (imec), Paola Favia (imec), Kris Vanstreels (imec), Marleen van der Veen (imec), Chen Wu (imec), Gayle Murdoch (imec), Nicolas Jourdan (imec), Antony Peter (imec), Bart Sor\'ee (imec, KU Leuven),  Bensu Tunca Alt\i{}nta\c{s} (imec), Nick Goossens (KU Leuven), Jef Vleugels (KU Leuven), Sean McMitchell (imec), Alfonso Sepulveda Marquez (imec), Sven Van Elshocht (imec), Christopher J. Wilson (imec), and J\"urgen B\"ommels (imec) for many fruitful discussions. Dawit Abdi (imec) and Odysseas Zografos (imec) are acknowledged for providing Fig.~\ref{IC_SRAM}b. The authors would also like to acknowledge the support provided by imec's pline and MCA department for the numerous experiments conducted on this topic over the past decade. This work has been supported by imec's industrial affiliate program on nano-interconnects.

\end{acknowledgments}

\section*{Author Declarations}

\subsection*{Conflict of Interest}

The authors have no conflicts to disclose. 

\subsection*{Author Contributions}

\textbf{Jean-Philippe Souli\'e:} Conceptualization (equal); Supervision (equal); Writing - original draft (lead); Writing - review \& editing (lead). \textbf{Kiroubanand Sankaran:} Conceptualization (equal); Writing - original draft (equal); Writing - review \& editing (equal). \textbf{Benoit Van Troeye:} Conceptualization (equal); Writing - original draft (equal); Writing - review \& editing (equal). \textbf{Alicja Le\'sniewska:} Conceptualization (equal); Writing - original draft (equal); Writing - review \& editing (equal). \textbf{Olalla Varela Pedreira:} Conceptualization (equal); Writing - original draft (equal); Writing - review \& editing (equal). \textbf{Herman Oprins:} Conceptualization (equal); Writing - original draft (equal); Writing - review \& editing (equal). \textbf{Gilles Delie:} Conceptualization (equal); Writing - original draft (equal); Writing - review \& editing (equal). \textbf{Claudia Fleischmann:} Conceptualization (equal); Project administration (equal); Supervision (equal); Writing - review \& editing (equal). \textbf{Lizzie Boakes:} Conceptualization (equal); Writing - original draft (equal); Writing - review \& editing (equal). \textbf{C\'edric Rolin:} Conceptualization (equal); Project administration (equal); Supervision (equal); Writing - review \& editing (equal). \textbf{Lars-\AA{}ke Ragnarsson:} Conceptualization (equal); Project administration (equal); Supervision (equal); Writing - review \& editing (equal). \textbf{Kristof Croes:} Conceptualization (equal); Project administration (equal); Supervision (equal); Writing - review \& editing (equal). \textbf{Seongho Park:} Conceptualization (equal); Funding acquisition (equal); Project administration (equal); Supervision (equal); Writing - review \& editing (equal). \textbf{Johan Swerts:} Conceptualization (equal); Funding acquisition (equal); Project administration (equal); Supervision (equal); Writing - review \& editing (equal). \textbf{Geoffrey Pourtois:} Conceptualization (equal); Funding acquisition (equal); Project administration (equal); Supervision (equal); Writing - review \& editing (equal). \textbf{Zsolt T\H{o}kei:} Conceptualization (equal); Funding acquisition (lead); Project administration (equal); Supervision (equal); Writing - review \& editing (equal). \textbf{Christoph Adelmann:}  Conceptualization (lead); Funding acquisition (equal); Project administration (equal); Supervision (equal); Writing - original draft (lead); Writing - review \& editing (lead). 

\section*{Data availability}

The data that supports the findings of this study are available within the article. 

\bibliographystyle{aip}
\bibliography{references.bib}

\end{document}